\documentclass[]{aa}

\usepackage{pgf}
\usepackage{graphicx}
\usepackage{txfonts}
\usepackage{natbib,twoopt}
\usepackage{hyperref} %% to avoid \citeads line fills
\usepackage{caption} 
\usepackage{subcaption}
\usepackage{ragged2e}
\usepackage{array}
\usepackage{rotating}
\newcolumntype{L}[1]{>{\raggedright\let\newline\\\arraybackslash\hspace{0pt}}m{#1}}
\newcolumntype{C}[1]{>{\centering\let\newline\\\arraybackslash\hspace{0pt}}m{#1}}
\newcolumntype{R}[1]{>{\raggedleft\let\newline\\\arraybackslash\hspace{0pt}}m{#1}}

\bibpunct{(}{)}{;}{a}{}{,}             %% natbib format for A&A and ApJ
\makeatletter
  \newcommandtwoopt{\citeads}[3][][]{\href{http://adsabs.harvard.edu/abs/#3}%
    {\def\hyper@linkstart##1##2{}%
     \let\hyper@linkend\@empty\citealp[#1][#2]{#3}}}
  \newcommandtwoopt{\citepads}[3][][]{\href{http://adsabs.harvard.edu/abs/#3}%
    {\def\hyper@linkstart##1##2{}%
     \let\hyper@linkend\@empty\citep[#1][#2]{#3}}}
  \newcommandtwoopt{\citetads}[3][][]{\href{http://adsabs.harvard.edu/abs/#3}%
    {\def\hyper@linkstart##1##2{}%
     \let\hyper@linkend\@empty\citet[#1][#2]{#3}}}
  \newcommandtwoopt{\citeyearads}[3][][]%
    {\href{http://adsabs.harvard.edu/abs/#3}
    {\def\hyper@linkstart##1##2{}%
     \let\hyper@linkend\@empty\citeyear[#1][#2]{#3}}}
\makeatother

\begin{document}

\title{MUSE library of stellar spectra}

\author{
Valentin D. Ivanov\inst{1}
\and
Lodovico Coccato\inst{1}
\and
Mark J. Neeser\inst{1}
\and
Fernando Selman\inst{4}
\and
Alessandro Pizzella\inst{2,3}
\and
Elena Dalla Bont\'a\inst{2,3}
\and
Enrico M. Corsini\inst{2,3}
\and
Lorenzo Morelli\inst{5}
}

\offprints{V. Ivanov, \email{vivanov@eso.org}}

\institute{
% 1
European Southern Observatory, Karl-Schwarzschild-Str. 2, 85748 Garching bei M\"unchen, Germany
\and 
% 2
Dipartimento di Fisica e Astronomia ``G. Galilei'', Università di Padova, Vicolo dell'Osservatorio 3, 35122, Padova, Italy
\and
% 3
INAF-Osservatorio Astronomico di Padova, Vicolo dell'Osservatorio 5, 35122, Padova, Italy
\and
% 4
European Southern Observatory, Ave. Alonso de C\'ordova 3107, Vitacura, Santiago, Chile
\and
% 5
Instituto de Astronom\'ia y Ciencias Planetarias Universidad de Atacama, Copiap\'o, Chile
}

\date{Received 2 November 1002 / Accepted 7 January 3003}

\abstract 
% context
{Empirical stellar spectral libraries have applications in both 
extragalactic and stellar studies, and they have an advantage 
over theoretical libraries because they naturally include all 
relevant chemical species and physical processes.
During recent years we see a stream of new high quality sets 
of spectra, but increasing the spectral resolution and widening 
the wavelength coverage means resorting to multi-order echelle 
spectrographs. Assembling the spectra from many pieces results 
in lower fidelity of their shapes.}
% aims
{We aim to offer the community a library of high signal-to-noise
spectra with reliable continuum shapes. Furthermore, the using 
an integral field unit (IFU) alleviates the issue of slit 
losses.}
% method
{Our library was build with the MUSE (Multi-Unit Spectroscopic 
Explorer) IFU instrument. We obtained spectra over nearly the 
entire visual band ($\lambda$$\sim$4800--9300\,\AA).}
% results
{We assembled a library of 35 high-quality MUSE spectra for a 
subset of the stars from the X-shooter Spectral Library. We 
verified the continuum shape of these spectra with synthetic 
broad band colors derived from the spectra. We also report some 
spectral indices from the Lick system, derived from the new 
observations.}
% conclusions
{We offer a high-fidelity set of stellar spectra that covers 
the Hertzsprung-Russell diagram. It can be useful for both 
extragalactic and stellar work and demonstrates that the IFUs 
are excellent tools for building reliable spectral libraries. 
}

\keywords{atlases -- stars:abundances -- stars:fundamental 
parameters -- stars:atmospheres -- galaxies:stellar content}
\authorrunning{V. Ivanov et al.}
\titlerunning{MUSE stellar library}

\maketitle

\section{Introduction}\label{sec:intro}

Empirical stellar spectral libraries are one of the most universal 
tools in modern astronomy. They have applications in both 
extragalactic and in stellar studies. The former include the 
modelling of unresolved stellar populations 
\citep[e.g.][]{2016A&A...589A..73R}, matching and removing continua 
to reveal weak emission lines \citep[e.g.][]{1998ApJ...505..639E}, 
usage as templates to measure the stellar line-of-sight velocity 
dispersions in galaxies 
\citep[][]{1977ApJ...212..326S,2015MNRAS.452....2K,2018MNRAS.480.3215J,2018A&A...612A..66M,2019A&A...623A..87N}. 
The stellar applications include measuring stellar parameters such 
as effective temperatures \citep[e.g. ][]{2015MNRAS.454.4054B} and 
surface gravities \citep[e.g.][]{2015ApJ...802L..10T} by template 
matching or indices, measuring radial velocities 
\citep[e.g.][]{2016MNRAS.456.4315S}, and verifying theoretical 
stellar models which sometime are not as good as one may expect. 
For example, 
\citet{2013MNRAS.435..952S} found discrepancies in the Balmer 
lines, suggesting that the theoretical spectral libraries may not 
be as reliable source of stellar spectra as the empirical ones.
The lists of applications given here are by far incomplete. 

We can add a number of open issues related to the libraries: the 
need to derive homogenious and self-consistent stellar parameters 
of the library stars -- right now the stellar parameters are 
typically assembled from multiple sources. This requires a two 
step process: first, derive global solutions of stellar parameters 
$T_{\rm eff}$/[Fe/H]/$\log g$ versus spectral indices, and then 
to invert these relations and to derive new uniform set of stellar 
parameters for all stars 
\citep[e.g.]{2016A&A...585A..64S,2019A&A...627A.138A}.
Another issue is to define optimal indices, most sensitive to one 
or another stellar parameter \citep[e.g.][]{2013A&A...549A.129C}. 
A particular problem related to galaxy models is the contribution 
of the AGB stars \citep[e.g.][]{2005MNRAS.362..799M}.

The most widely used theoretical libraries today are the BaSeL 
\citep{1992IAUS..149..225K,1997A&AS..125..229L,1998A&AS..130...65L,2002A&A...381..524W}
and the PHOENIX \citep{1999ApJ...525..871H,2012RSPTA.370.2765A,2013A&A...553A...6H},
but there have been 
problems with the treatment of molecules, as shown early on by 
\citet{1997A&A...318..841C}, that occasionally lead to poorly 
predicted broad band colors. Among the empirical libraries the
work of \citet{1998PASP..110..863P} was the most widely used. It 
includes 131 flux calibrated stars, but for the vast majority of 
them the resolving power was below $R$=1000, which is relatively 
low even for extragalactic applications where the intrinsic 
velocity dispersion of galaxies require $R$$\sim$2000 or higher. 
Other sets of spectra with better quality have become available: 
ELODIE \citep{1998A&AS..133..221S,2001A&A...369.1048P,2004A&A...425..881L}, 
STELIB \citep{2003A&A...402..433L}, 
Indo-US \citep{2004ApJS..152..251V}, 
MILES \citep{2006MNRAS.371..703S}, and 
CaT \citep{2001MNRAS.326..959C, 2007MNRAS.374..664C}.
More recently, single order library with a large number of stars 
was reported by \citet{2018arXiv181202745Y}, but it has been 
obtained with a 3\,arcsec fibers of the SDSS spectrograph 
\citep{2017AJ....154...28B} and therefore does not avoid 
completely the slit loss problem. \citet{2011MNRAS.418.2785M} 
incorporated some of the libraries listed here in a comprehensive 
stellar population model at high spectral resolution. 

The X-shooter Spectral Library \citep[XSL;][]{2014A&A...565A.117C} 
is the latest and most comprehensive effort in this direction. At
this time only the optical spectra are available. At this 
time only the Data Release 1 was available. It contains 237 
stars and when completed it will cover 0.3--2.5\,$\mu$m range at 
a resolving power of $R$$\sim$7000--11000. The XSL is a good 
example of the problems that increasing resolution and multi-order 
cross-dispersed spectrographs bring in: the synthetic broad band 
optical ($UBV$) colors agree poorly with the observed colors 
from the Bright Star Catalog \citep[on average at $\sim$7\,\% 
level, see Table\,5 and Fig.\,26 in][]{2014A&A...565A.117C}. The 
differences are partially related to pulsating variable stars 
having been observed in different phases. Slit losses are another 
issue; 
for many stars that is caused by the lack or the poor quality wide 
slit observations. Despite these problems, the narrow features in 
the XSL spectra are self-consistent, e.g. observations in different 
orders agree well \citep[see Fig.\,8 in][]{2014A&A...565A.117C}, 
and there is a good agreement between features and theoretical 
models and other empirical libraries \citep[for a comparison with 
the UVES-POP see Figs.\,31--34 in][]{2014A&A...565A.117C}.

In other words, we are facing again a familiar problem: the old 
theoretical libraries used to predict colors inconsistent with 
the observations; nowdays, the newest empirical libraries do the 
same, despite -- or because -- of the excellent quality of the new 
data, that made it more apparent.
To address this issue we embarked on a project to build a {\it 
slitloss-less} empirical spectral library with the MUSE 
\citep[Multi-Unit Spectroscopic Explorer;][]{2010SPIE.7735E..08B} 
integral field unit, spanning all major sequences on the 
Hertzsprung-Russell diagram, with the specific goal of adjusting 
and verifying the shapes of the spectra in other libraries, both 
theoretical and empirical. The final product are spectra, suitable 
for galactic modeling, stellar classification, and other 
applications. Here we report the first subset of 35 MUSE stellar 
spectra.

The next two sections describe the sample and the data, respectively. 
Section\,\ref{sec:analysis} presents the analysis of our spectra and 
Sec.\,\ref{sec:summary} summarizes this work.

\section{Sample}\label{sec:sample}

Our initial sample numbered 33 targets selected among the XSL 
stars\footnote{\url{http://xsl.u-strasbg.fr/}}. We aimed to 
populate the Hertzsprung-Russell diagram as homogeneously as 
possible with $\sim$3-6 bright stars per spectral type, ensuring a 
high signal-to-noise S/N$>$70--200 per spectral type, except for 
the O-type where only a single star was available.

Spectra of two additional stars were obtained: HD\,193256 and 
HD\,193281B. They serendipitously fell inside the field of view 
during the observations of the project target HD\,193281A. An IFU 
campaign covering the entire XSL is planned, but we made sure to 
select stars over various spectral types, making this trimmed-down 
library adequate for some applications, such as stellar 
classification and templates fitting of galaxy spectra. 

The SIMBAD spectral types as listed in \citet{2014A&A...565A.117C}, 
and complemented for the two extra targets, together with effective 
temperatures $T_{\rm eff}$, surface gravities {\it log g} and 
metalicities [Fe/H] collected from the literature, if available, 
are listed in Table\,\ref{tab:params} and shown in 
Fig.\,1\footnote{Stellar parameters from XSL also became available 
after the submission of this paper: \citet{2019A&A...627A.138A}}.
The covered range of $T_{\rm eff}$ is 2600-33000\,K, of {\it log g}: 
0.6-4.5 and of [Fe/H]: from $-$1.22 to 0.55, as far as the stellar 
parameters as known. In case multiple literature sources with 
equal quality were available for a certain parameter, we adopted the 
average value and if a given source had significantly smaller errors 
than the others, we adopted the value from that source.

\begin{figure}[!htb]
\centering
\includegraphics[width=7cm]{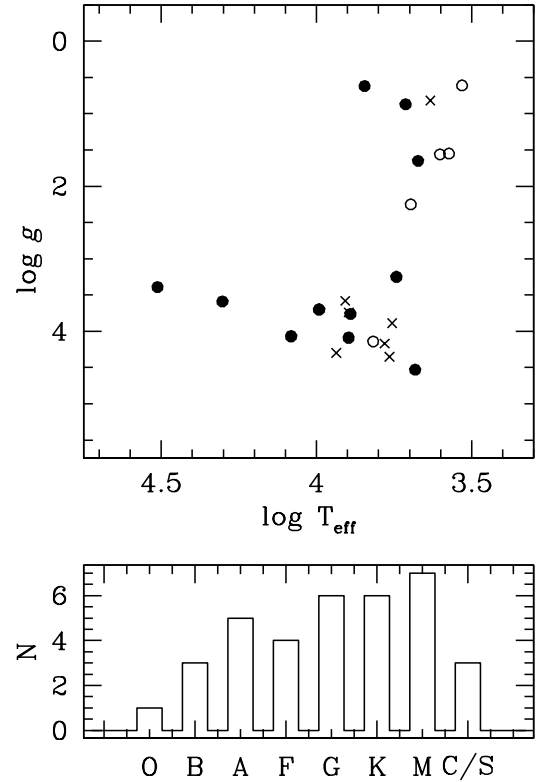}\\
\caption{Properties of the stars in our sample.
{\it Top}: Surface gravity {\it log g} versus effective temperature 
$T_{\rm eff}$ for stars with [Fe/H]$\leq$$-$0.5\,dex (crosses), 
$-$0.5$<$[Fe/H]$<$0.0\,dex (open circles) and [Fe/H]$\geq$0.0\,dex 
(solid dots).
{\it Bottom}: Distributions of the stars by spectral type.
}\label{fig:params}
\end{figure}

\begin{table*}
\caption{Physical parameters of the program stars. The columns 
contain: (1) object ID (asterisks mark non-XSL objects); (2) 
SIMBAD spectral type; (3-4) radial velocity and reference; (4-8) 
effective temperature, surface gravity, iron abundance and 
reference. Our estimated spectral type and effective temperature 
for HD\,193281B are also listed.}\label{tab:params}
\begin{center}
\begin{small}
\begin{tabular}{@{ }l@{ }c@{}c@{}l@{ }c@{}c@{}c@{}l@{ }}
\hline\hline
IDs & Sp.\,Type & ~~~V$_{\rm rad}$,\,km\,s$^{-1}$~~~~ & Reference~~~~~~~~~~~~~~~~~~~~~~~~~ & ~~~~~~~$T_{\rm eff}$,\,K~~~~~~ & ~~~~log\,$g$~~~~~ & ~~~~~~[Fe/H]~~~~~~ & Reference\\
(1) & (2) & (3) & (4) & (5) & (6) & (7) & (8) \\
\hline
HD\,057060                          & O7e...  &        20.0$\pm$1.7 & \protect{\citet{2004A&A...424..727P}} & 32508$\pm$1928 & 3.39$\pm$0.26 &     0.24$\pm$0.14 & \protect{\citet{2012A&A...538A.143K}}  \\ % 1
                        &         &                     &                                       & 33215$\pm$2674 & 3.28$\pm$0.16 &  $-$0.03$\pm$0.20 & \protect{\citet{2011A&A...531A.165P}}  \\ % 1
HD\,064332              & S       &      $-$1.3$\pm$0.5 & \protect{\citet{2006AstL...32..759G}} &  3399$\pm$44   & 0.61$\pm$0.40 &  $-$0.04$\pm$0.18 & \protect{\citet{2011A&A...531A.165P}}  \\ % 2
HD\,067507              & CNv...  &           23$\pm$10 & \protect{\citet{1953GCRV..C......0W}} &  2680          & ...           &  ...              & \protect{\citet{2002A&A...390..967B}}  \\ % 3 RU\,Pup
HD\,085405              & C       &        3.50$\pm$1.6 & \protect{\citet{2006AstL...32..759G}} &  2769          & ...           &  $-$0.10          & \protect{\citet{2016A&A...591A.118S}}  \\ % 4 Y\,Hya
                        &         &                     &                                       &  2645          & ...           &  ...              & \protect{\citet{2002A&A...390..967B}}  \\ % 4 Y\,Hya
HD\,096446              & B2IIIp  &         6.1$\pm$0.8 & \protect{\citet{2006AstL...32..759G}} & 20086$\pm$530  & 3.59$\pm$0.08 &     0.06$\pm$0.04 & \protect{\citet{2012A&A...538A.143K}}  \\ % 5
HD\,099648              & G8Iab   &    $-$8.82$\pm$0.19 & \protect{\citet{2018yCat.1345....0G}} &  4970$\pm$75   & 2.25$\pm$0.43 &  $-$0.01$\pm$0.15 & \protect{\citet{2012A&A...538A.143K}}  \\ % 6
                        &         &                     &                                       &  4977$\pm$49   & 2.24$\pm$0.12 &  $-$0.03$\pm$0.06 & \protect{\citet{2011A&A...531A.165P}}  \\ % 6
HD\,099998              & K3.5III &      18.43$\pm$0.37 & \protect{\citet{2018yCat.1345....0G}} &  4001$\pm$32   & 1.56$\pm$0.20 &  $-$0.24$\pm$0.07 & \protect{\citet{2011A&A...531A.165P}}  \\ % 7 BS\,4432
HD\,100733              & M3III   &      21.07$\pm$0.29 & \protect{\citet{2018yCat.1345....0G}} &  3530          & ...           &  ...              & \protect{\citet{2003AJ....125..359W}}  \\ % 8 BS\,4463
HD\,306799              & M0Iab   &   $-$16.38$\pm$0.19 & \protect{\citet{2008A&A...485..303M}} &  3650          & ...           &  ...              & \protect{\citet{2003AJ....125..359W}}  \\ % 9 CD-60\,3636
HD\,101712              & M3Iab   &    $-$0.70$\pm$1.23 & \protect{\citet{2008A&A...485..303M}} &  3200          & ...           &  ...              & \protect{\citet{2003AJ....125..359W}}  \\ % 10
HD\,102212              & M1III   &      50.28$\pm$0.09 & \protect{\citet{2009A&A...498..627F}} &  3738$\pm$6    & 1.55$\pm$0.10 &  $-$0.41$\pm$0.05 & \protect{\citet{2012A&A...538A.143K}}  \\ % 11
HD\,114960              & K5III   &       7.35$\pm$0.16 & \protect{\citet{2018yCat.1345....0G}} &  4000          & ...           &  ...              & \protect{\citet{2003AJ....125..359W}}  \\ % 12
IRAS\,15060+0947        & M9III   &      $-$8.2$\pm$2.6 & \protect{\citet{2015A&A...582A..68E}} &  3281          & ...           &  ...              & \protect{\citet{2018A&A...616A...1G}}  \\ % 13
HD\,147550              & B9V     &     $-$24.1$\pm$0.9 & \protect{\citet{2006AstL...32..759G}} &  9830$\pm$279  & 3.70$\pm$0.66 &  $-$0.38$\pm$0.11 & \protect{\citet{2012A&A...538A.143K}}  \\ % 14
HD\,160365              & F6III   &       8.14$\pm$2.31 & \protect{\citet{2008AJ....135..209M}} &  6009          & ...           &  ...              & \protect{\citet{2016A&A...591A.118S}}  \\ % 15
HD\,160346              & K3V     &    17.856$\pm$0.784 & \protect{\citet{2017AJ....153...75K}} &  4808$\pm$65   & 4.53$\pm$0.22 &     0.03$\pm$0.10 & \protect{\citet{2012A&A...538A.143K}}  \\ % 16
HD\,163810              & G3V     &     185.99$\pm$0.22 & \protect{\citet{2002AJ....124.1144L}} &  5818$\pm$15   & 4.35$\pm$0.06 &  $-$1.20$\pm$0.04 & \protect{\citet{2012A&A...538A.143K}}  \\ % 17
HD\,164257              & A0      &         5.5$\pm$0.9 & \protect{\citet{2006AstL...32..759G}} &  9792$\pm$691  & 3.70$\pm$2.11 &     0.41$\pm$0.30 & \protect{\citet{2012A&A...538A.143K}}  \\ % 18
\lbrack B86\rbrack\,133 & M4      &            44$\pm$5 & this work                             &  4637          & ...           &  $-$0.21          & \protect{\citet{2018A&A...616A...1G}}, \\ % 19
                        &         &                     &                                       &  2645          & ...           &  ...              & \protect{\citet{2004ApJS..151..387I}}  \\ % 19
HD\,167278              & F2      &     $-$14.7$\pm$0.9 & \protect{\citet{2006AstL...32..759G}} &  6563$\pm$18   & 4.14$\pm$0.08 &  $-$0.21$\pm$0.04 & \protect{\citet{2012A&A...538A.143K}}  \\ % 20
HD\,170820              & K0III   &       2.84$\pm$0.06 & \protect{\citet{2008A&A...485..303M}} &  4707$\pm$57   & 1.65$\pm$0.13 &     0.17          & \protect{\citet{2011A&A...531A.165P}}  \\ % 21
HD\,172230              & A5      &     $-$36.8$\pm$0.8 & \protect{\citet{2006AstL...32..759G}} &  7772$\pm$102  & 3.76$\pm$0.44 &     0.55$\pm$0.14 & \protect{\citet{2012A&A...538A.143K}}  \\ % 22
HD\,173158              & K0      &      14.06$\pm$0.32 & \protect{\citet{2018yCat.1345....0G}} &  5164$\pm$121  & 0.87$\pm$0.43 &     0.04$\pm$0.20 & \protect{\citet{2012A&A...538A.143K}}  \\ % 23
HD\,174966              & A3      &         5.6$\pm$0.9 & \protect{\citet{2018yCat.1345....0G}} &  7874$\pm$57   & 4.09$\pm$0.16 &     0.03$\pm$0.10 & \protect{\citet{2012A&A...538A.143K}}  \\ % 24
HD\,175640              & B9III   &     $-$26.0$\pm$4.3 & \protect{\citet{2006AstL...32..759G}} & 12067$\pm$326  & 4.07$\pm$0.55 &     0.22$\pm$0.18 & \protect{\citet{2012A&A...538A.143K}}  \\ % 25
                        &         &                     &                                       & 12077$\pm$453  & 3.94$\pm$0.21 &     0.17$\pm$0.15 & \protect{\citet{2011A&A...531A.165P}}  \\ % 25
HD\,179821              & G5Ia    &      81.78$\pm$3.71 & \protect{\citet{2018yCat.1345....0G}} &  6997          & 0.62          &     0.44          & \protect{\citet{2016A&A...591A.118S}}  \\ % 26
                        &         &                     &                                       &  7107          & 1.00          &     0.45          & \protect{\citet{2016A&A...591A.118S}}  \\ % 26
HD\,232078              & K3IIp   &  $-$388.34$\pm$0.27 & \protect{\citet{2008A&A...480...91S}} &  4295$\pm$48   & 0.82$\pm$0.27 &  $-$1.08$\pm$0.11 & \protect{\citet{2012A&A...538A.143K}}  \\ % 27
                        &         &                     &                                       &  4014$\pm$48   & 0.81$\pm$0.20 &  $-$1.22$\pm$0.11 & \protect{\citet{2011A&A...531A.165P}}  \\ % 27
HD\,193256$^*$          & A8Vn... &             6$\pm$2 & this work                             &  7860          & 3.74          &  $-$0.95          & \protect{\citet{2016A&A...591A.118S}}  \\ % 28 https://ui.adsabs.harvard.edu/abs/2017AJ....154...31G/abstract
HD\,193281A             & A2III   &         0.3$\pm$0.5 & \protect{\citet{2006AstL...32..759G}} &  8623$\pm$345  & 4.30$\pm$0.33 &  $-$0.68$\pm$0.28 & \protect{\citet{2012A&A...538A.143K}}  \\ % 29
                        &         &                     &                                       &  8597$\pm$218  & 4.11$\pm$0.14 &  $-$0.37$\pm$0.13 & \protect{\citet{2011A&A...531A.165P}}  \\ % 29
HD\,193281B$^*$         & F5:V:   &   $-$43.13$\pm$0.97 & \protect{\citet{2018yCat.1345....0G}} &  8080          & 3.58          &  $-$1.00          & \protect{\citet{2016A&A...591A.118S}}  \\ % 30 https://ui.adsabs.harvard.edu/abs/1984ApJS...55..657C/abstract
                        &         &                     &                                       &  8080          & 3.58          &  $-$1.00          & \protect{\citet{2016A&A...591A.118S}}  \\ % 30
                        &         &                     &                                       &  8414          & ...           &  ...              & \protect{\citet{2016A&A...591A.118S}}  \\ % 30
                        & K2III   &                     &                                       &  4354$\pm$57   & ...           &  ...              & this work                              \\ % 30
HD\,193896              & G5IIIa  &   $-$15.23$\pm$0.18 & \protect{\citet{2018yCat.1345....0G}} &  4900          & ...           &  ...              & \protect{\citet{2003AJ....125..359W}}  \\ % 31
HD\,196892              & F6V     & $-$34.498$\pm$0.004 & \protect{\citet{2011A&A...526A.112S}} &  6028$\pm$22   & 4.17$\pm$0.10 &  $-$0.99$\pm$0.07 & \protect{\citet{2012A&A...538A.143K}}  \\ % 32
HD\,200081              & G0      &       7.67$\pm$0.27 & \protect{\citet{2008A&A...480...91S}} &  5526$\pm$71   & 3.25$\pm$0.43 &     0.02$\pm$0.12 & \protect{\citet{2012A&A...538A.143K}}  \\ % 33
HD\,204155              & G5      &   $-$84.60$\pm$0.16 & \protect{\citet{2002AJ....124.1144L}} &  5704$\pm$28   & 3.89$\pm$0.16 &  $-$0.70$\pm$0.07 & \protect{\citet{2012A&A...538A.143K}}  \\ % 34
                        &         &                     &                                       &  5718$\pm$56   & 3.93$\pm$0.11 &  $-$0.69$\pm$0.06 & \protect{\citet{2011A&A...531A.165P}}  \\ % 34
HD\,209290              & M0.5V   &    18.144$\pm$0.069 & \protect{\citet{2013A&A...552A..64S}} &  4031          & ...           &  $-$0.06          & \protect{\citet{2006ApJ...638.1004A}}  \\ % 35
\hline
\end{tabular}
\end{small}
\end{center}
\end{table*}

\section{Observations and Data Reduction}\label{sec:reduction}

The spectra were obtained with MUSE at the European Souther 
Observatory (ESO) Very Large Telescope, Unit Telescope 4, on 
Cerro Paranal, Chile. 
Table\,\ref{tab:obs_log} gives the observing log. We obtained 
six exposures for each target, except for HD\,204155 which was 
observed 12 times. To maximize the data yield most of the data 
were obtained under non-photometric conditions, so the absolute 
flux calibration is uncertain, but the ``true'' intrinsic shape 
is preserved, because there is no ``stitching'' of multiple 
orders and no variable slit losses due to atmospheric refraction.
We placed the science targets at the same spaxels as the 
spectrophotometric standards, to minimize the instrument 
systematics that might arise from residual spaxel-to-spaxel 
variations.

The data reduction was performed with the ESO MUSE pipeline (ver. 
2.6) within the ESO 
Reflex\footnote{\url{https://www.eso.org/sci/software/esoreflex/}} 
environment \citep{2013A&A...559A..96F}. 
The 1-dimensional spectra were extracted within a circular aperture 
with a radius of 6\,arcsec. This number was selected after some 
experiments with apertures of difference sizes, to guarantee that 
``aperture'' losses will lead to a change in the overall slope of 
the spectra $<$1\,\% from the blue to the red end. The sky emission 
was estimated within an annulus of an inner radius 7\,arcsec and a 
width of 4\,arcsec. This step of the analysis was performed with an
IRAF\footnote{IRAF is distributed by the NOAO, which is operated by 
the AURA Inc., under contract to the NSF.}/PyRAF tool 
\citep{1986SPIE..627..733T,1993ASPC...52..173T,2012ascl.soft07011S}. 

Three stars were treated differently. 
For [B86]\,133 we reduced the extraction aperture radius to 
4\,arcsec (keeping the sky annulus the same as for the majority 
of the targets) to avoid contamination from nearby sources -- 
because the object is located in a crowded Milky Way bulge field.
HD\,193256 is close to the edge of the MUSE field of view, and 
the extraction apertures had to be smaller, with a radius 
4.6\,arcsec, the sky annulus had an inner radius of 4.6\,arcsec 
and a width of 2\,arcsec. 
HD\,193281 is a binary with $\sim$3.8\,arcsec separation and the 
components cross-contaminate each other. To separate the two 
spectra we first extracted a combined spectrum of the two stars 
together with the same aperture and annulus as for the bulk of 
the stars. 
Next, we rotated each plane of the data cube by 180\degr\ around 
the centre of the primary and subtracted the rotated plane from 
the original non-rotated 
plane, to remove the contribution of the primary at 
the location of the secondary. Then, we extracted the spectrum of 
the secondary with an aperture with a radius of 1.2\,arcsec and a 
sky annulus with an inner radius of 1.8\,arcsec and a width of 
4\,arcsec. Finally, we decontaminated the spectrum of the primary 
by subtracting the spectrum of the secondary from the combined 
spectrum of the binary.

Experiments with apertures of different sizes indicated that the
continuum shape of [B86]\,133 still changed at $<$1\% level across 
the entire wavelength range, despite the narrower extraction 
aperture. The spectra of the two other objects are less reliable 
and in the case of HD\,193281B a change in the radius of a few 
spaxels (0.2\,arcsec) leads to a flux change of $\sim$3\,\% over
the entire wavelength range. However, the spectrum of HD\,193281A 
is still stable at $<$1\,\% because the secondary contributes 
$\sim$1 and $\sim$11\,\% to the total flux at the blue and at the 
red ends of the spectrum, respectively, so this $\sim$3\,\% 
uncertainty is reduced by a factors of $\sim$100 and $\sim$9, 
respectively, and the spectrum of HD\,193281A can be considered 
reliable by to our criterion for $<$1\% stability across the 
entire spectral range.

The telluric features were removed by running {\it molecfit} ver. 
1.5.7 \citep{2015A&A...576A..77S,2015A&A...576A..78K} separately 
on each of the six (12 for HD\,204155) target spectra themselves.
The agreement of individual solutions is excellent: typically the 
fits yield a precipitable water estimate identical to within 
$<$0.1\,mm.

\begin{figure*}[!htb]
\centering
\includegraphics[height=15cm]{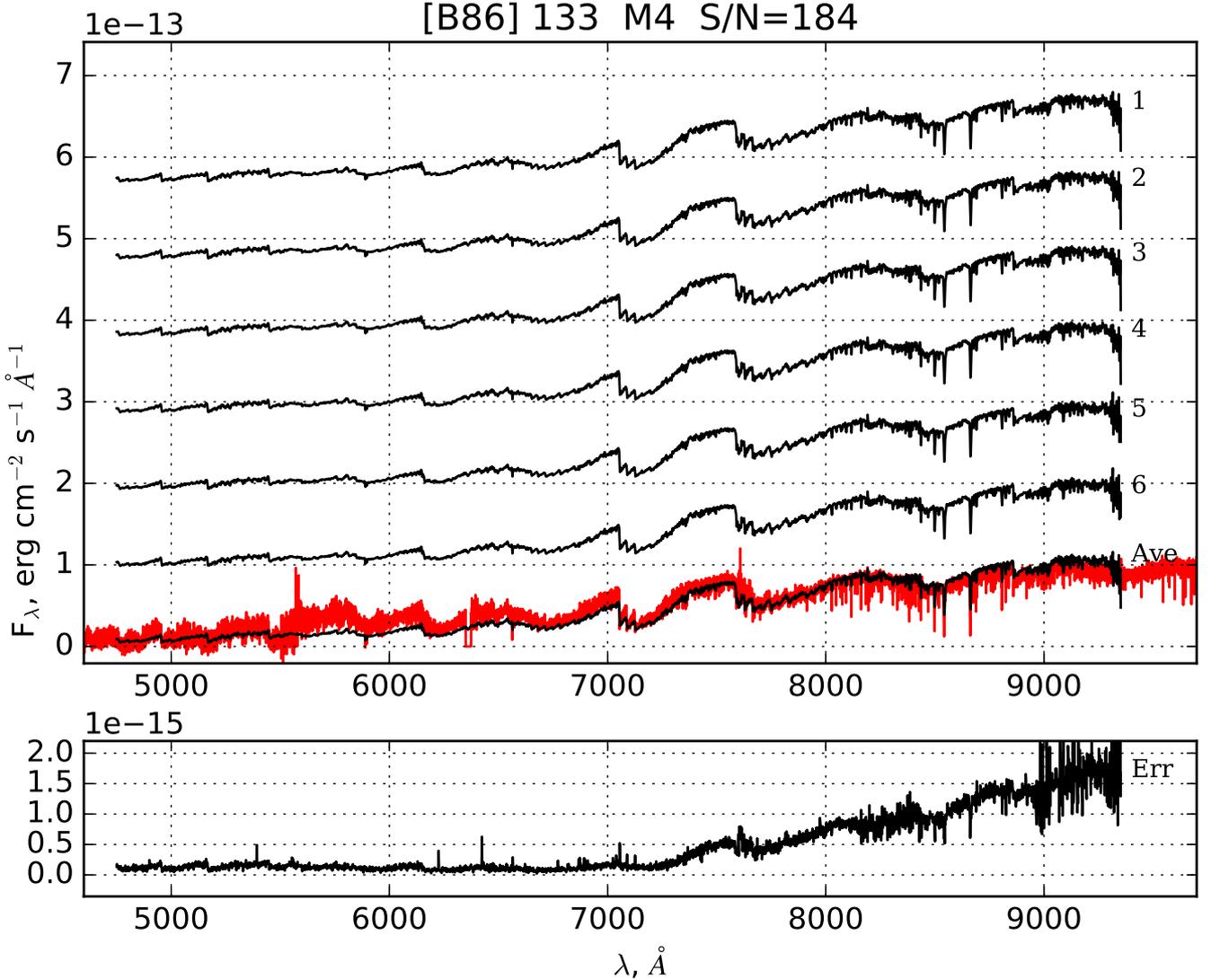} 
\caption{An example of the MUSE spectra (black line) of \lbrack 
B86\rbrack\,133 and of the corresponding XSL DR1 spectrum (red line;
normalized to match the MUSE spectrum flux). The plot title lists the
spectral type and the measured median S/N per resolution element over 
the entire spectrum.
The upper sub-panels show the spectra extracted from each individual 
exposure (shifted up for clarity) and the average spectra of the 
object at its true flux level. 
The bottom sub-panels show the standard deviation of the average 
spectrum.
The spectra of the other sample stars are presented in the electronic 
edition only (Fig.\,\ref{fig:spectra_full_sample}).}\label{fig:spectra_example}
\end{figure*}

The final spectrum for each target is the average of the 1-D spectra 
derived from the six individual observations, and the error is the 
r.m.s. of that averaging. A example of the data products is 
plotted in Fig.\,\ref{fig:spectra_example}. The complete sample is 
shown in Fig.\,\ref{fig:spectra_full_sample}. All final spectra are 
given in Table\,\ref{table:full_spectra} and are available in 
machine readable form at the journal's website.

\section{Analysis}\label{sec:analysis}

\begin{figure*}[!htb]
\centering
\includegraphics[height=9cm]{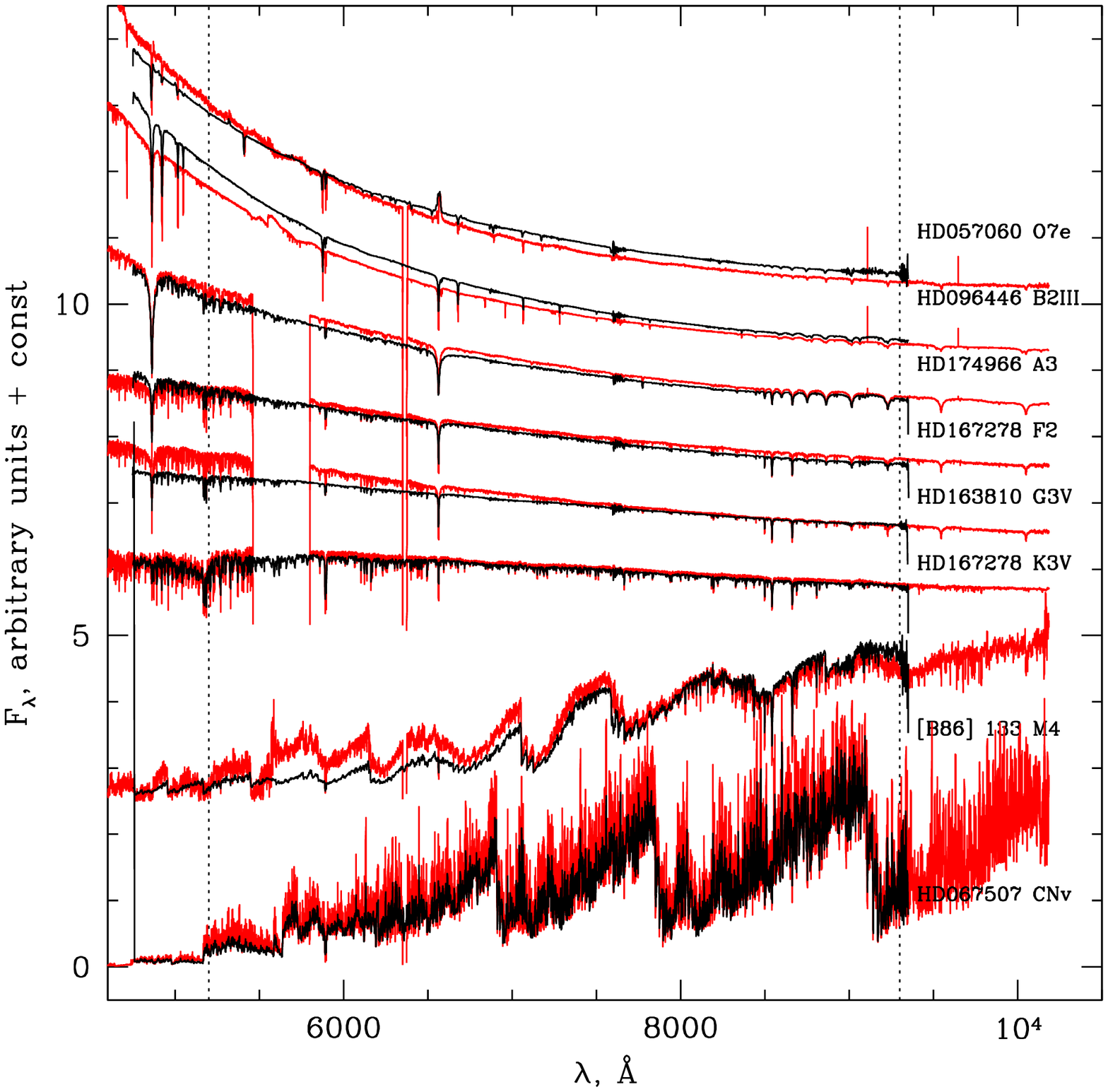} \includegraphics[height=9cm]{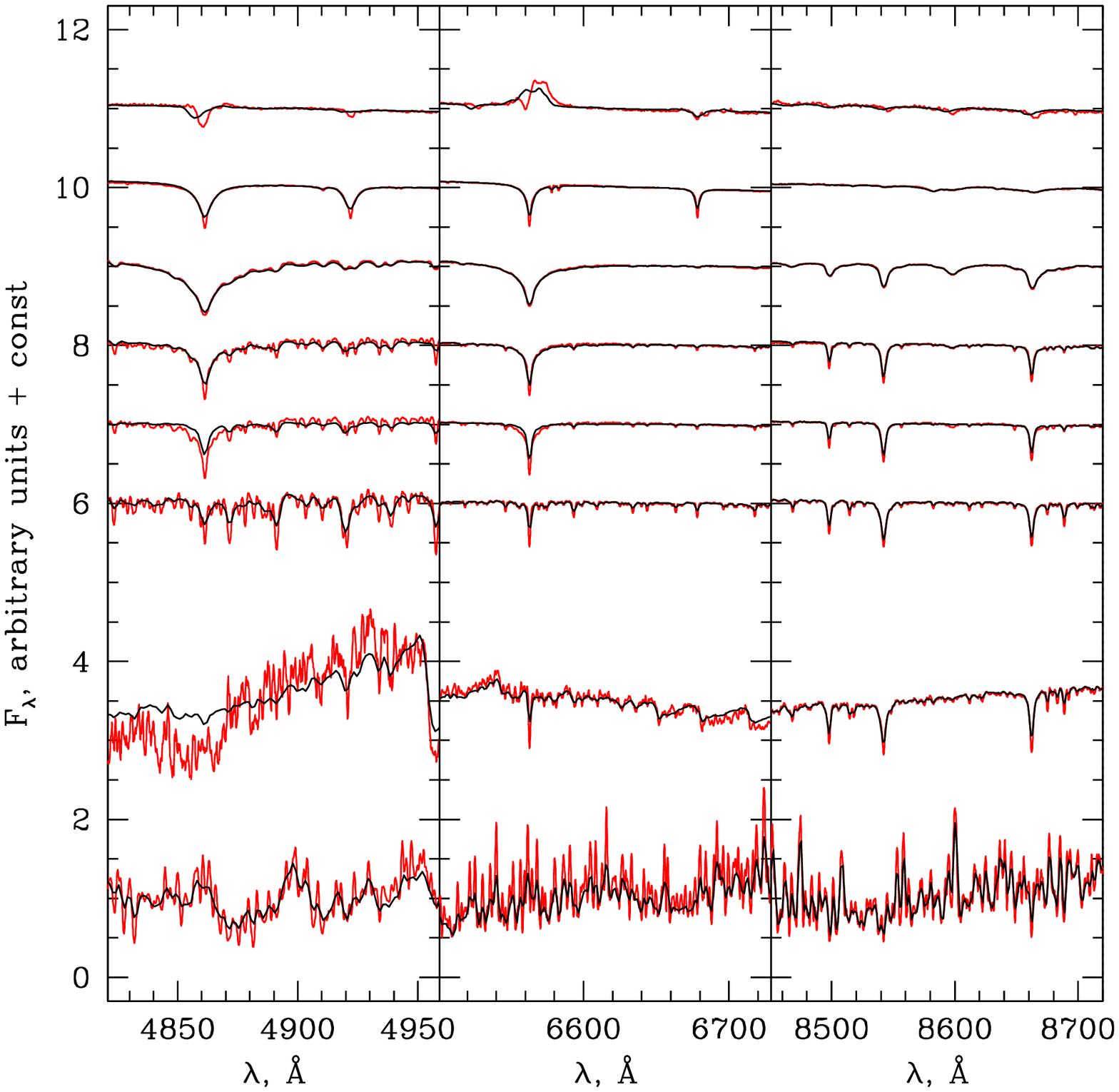} \\
\caption{Comparison of a subset of our MUSE spectra (black lines) 
with the XSL spectra (red lines; boxcar smoothed over 8 pixels). 
The spectra 
are normalized to unity between the two vertical dotted lines 
shown on the left panel, and shifted vertically for display 
purposes. The left panel shows the entire MUSE spectral range, 
the right panel zooms onto the H$\beta$, H$\alpha$, and Ca triplet 
wavelength ranges (left to right). No radial velocity corrections 
are applied.}\label{fig:MUSE_XSL}
\end{figure*}

\begin{figure*}[]
\centering
\includegraphics[width=15cm]{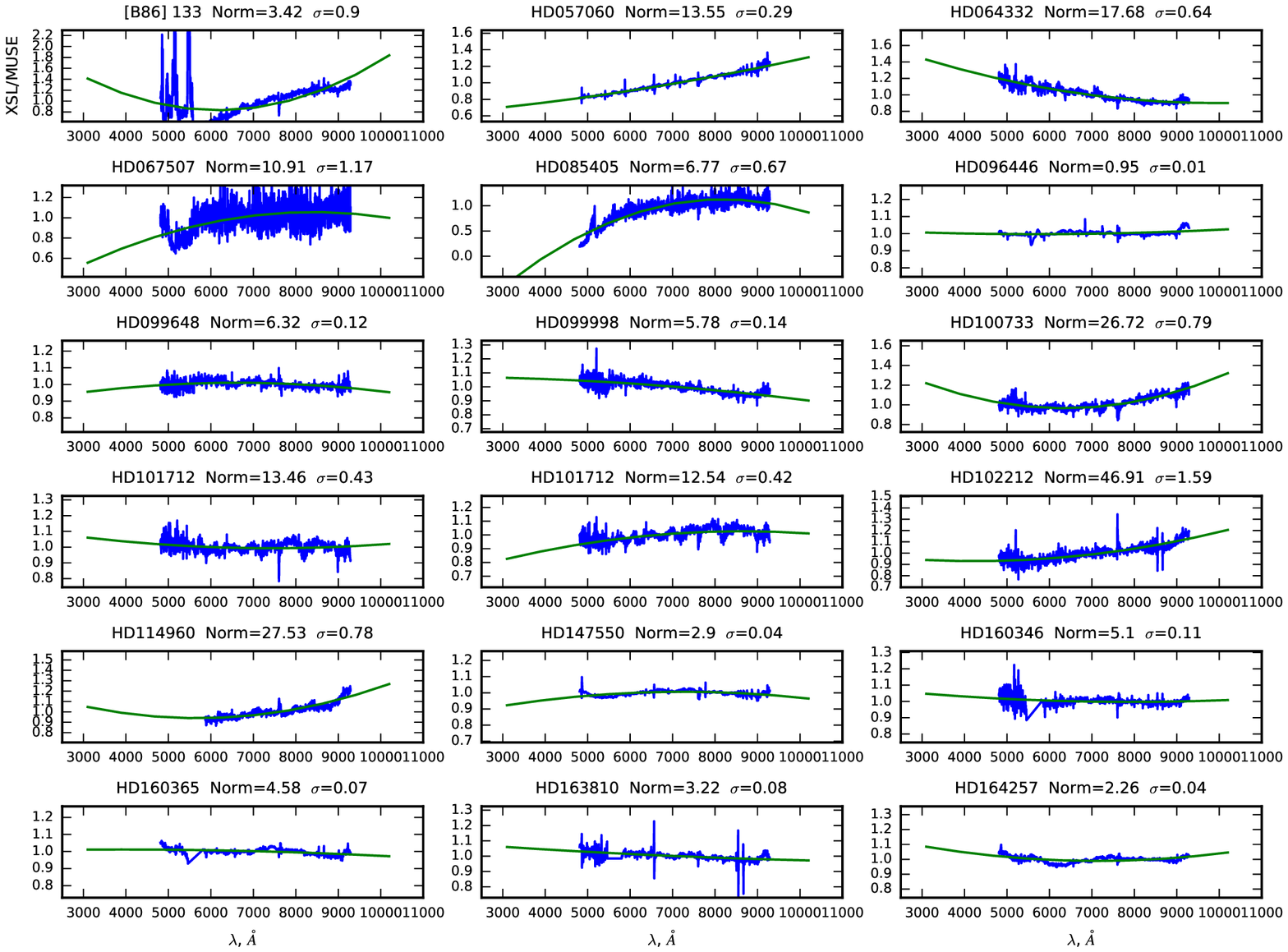} \\
\includegraphics[width=15cm]{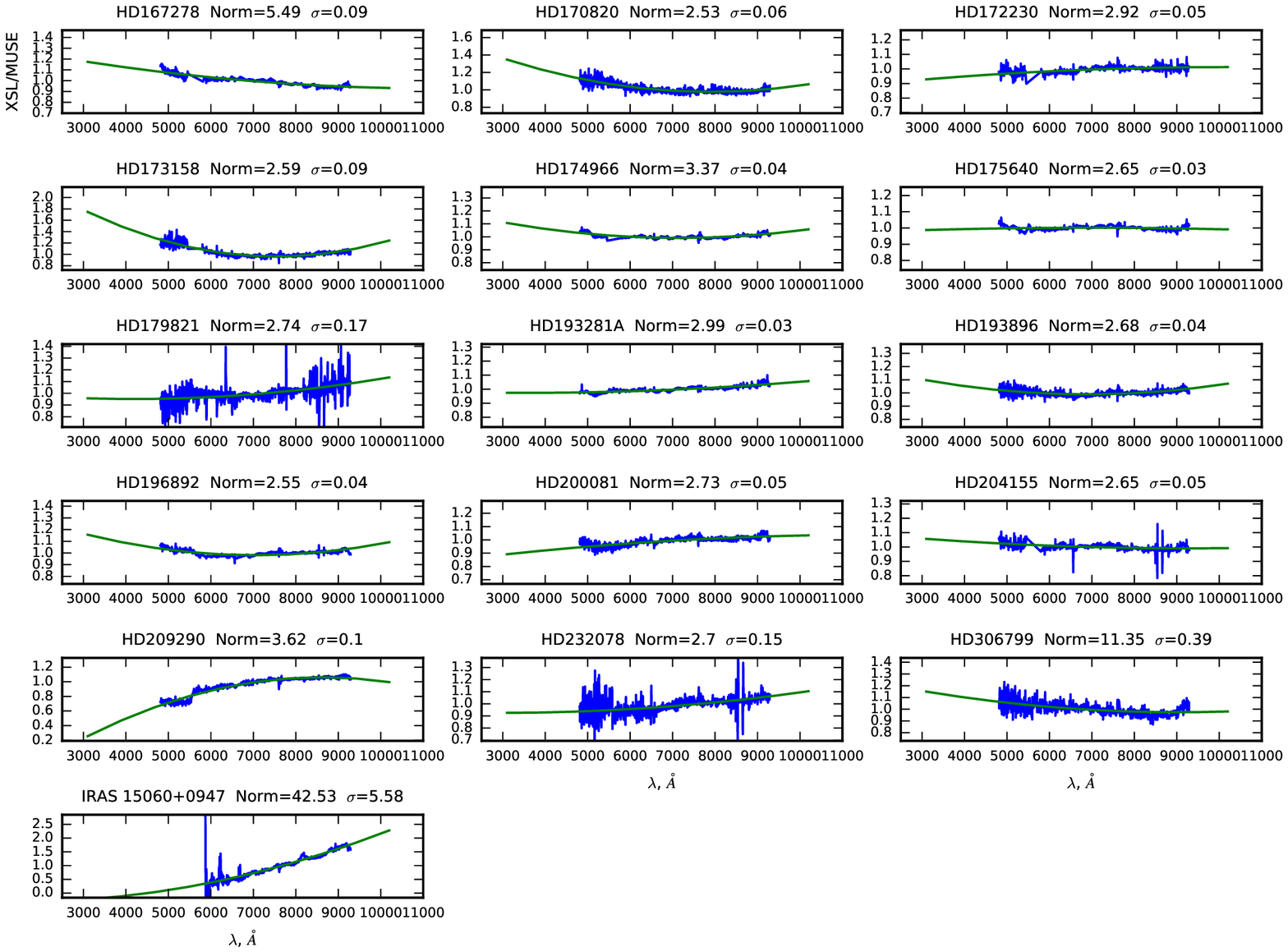} \\
\caption{Ratios of the XSL spectra to our MUSE spectra (blue; 
covers only the MUSE wavelength range), normalized to unity 
and median smoothed for display purposes with a 5-element 
wide median filter. Two ratios are show for HD\,101712 -- 
for the two XSL spectra of this star.
A second order polynomial fits spanning the wavelength of XSL is 
also shown in blue. The labels on the top of each panel contain 
the name of the object, the normalization factor that indicates 
the flux ratio of the independently flux-calibrated MUSE and 
XSL spectra, and a standard deviation of the fits residuals. 
The coefficients of polynomial fits are listed in 
Table\,\ref{tab:ratio_fits}.}\label{fig:spectra_ratios}
% \addtocounter{figure}{-1}
\end{figure*}

A direct comparison of the MUSE and XSL spectra for eight randomly 
selected stars across the spectral type sequence is shown with some 
zoomed-in spectral regions in Fig.\,\ref{fig:MUSE_XSL} (for the 
rest of out spectra see Figs.\,\ref{fig:spectra_example} and 
\ref{fig:spectra_full_sample}). Notably, the XSL spectra used
the continuum shape from a 5\,arcsec wide-slit observaitons.
In most cases the agreement on a 
scale of a few hundred pixels--in other words, within the same 
X-shooter order--is excellent. However, on wider scale we find 
deviations between the XSL and MUSE spectra, as can be seen in 
Fig.\,\ref{fig:spectra_ratios}. The exceptions are usually late 
type stars -- \lbrack B86\rbrack\,133 and IRAS\,15060+0947 are 
examples -- where the low signal-to-noise in the blue ($\sim$10 
or bellow) and the variability that only occurs with extremely 
red stars may account for the problem. 
Furthermore, the ratios of many spectra show gradual change, 
despite of their apparently high signal-to-noise: HD\,147550 and 
HD\,167278 are examples where the amplitude of the ratio within 
the MUSE wavelength range reaches 10-15\,\%. We fitted to the 
ratios second order polynomials and extrapolated them over the 
full wavelength range covered by the XSL library to demonstrate 
that if these trends hold, the overall peak-to-peak flux 
differences can easily rich $\sim$20\,\%, so the overall 
continuum of the cross-dispersed spectra is somewhat ill-defined. 
The coefficients of the polynomial fits are listed in 
Table\,\ref{tab:ratio_fits}. and can be used to correct the shape 
of the XSL spectra. We are far from critisizing 
\citet{2014A&A...565A.117C} for the quality of their data 
reduction, rather we point here that the high signal-to-noise 
observations show how difficult it is to process cross-dispersed 
spectra. Indeed, problems that may not be obvious with poor 
quality data become apparent for signal-to-noise if 100-200.

\begin{figure*}[!htb]
\centering
\includegraphics[width=16cm]{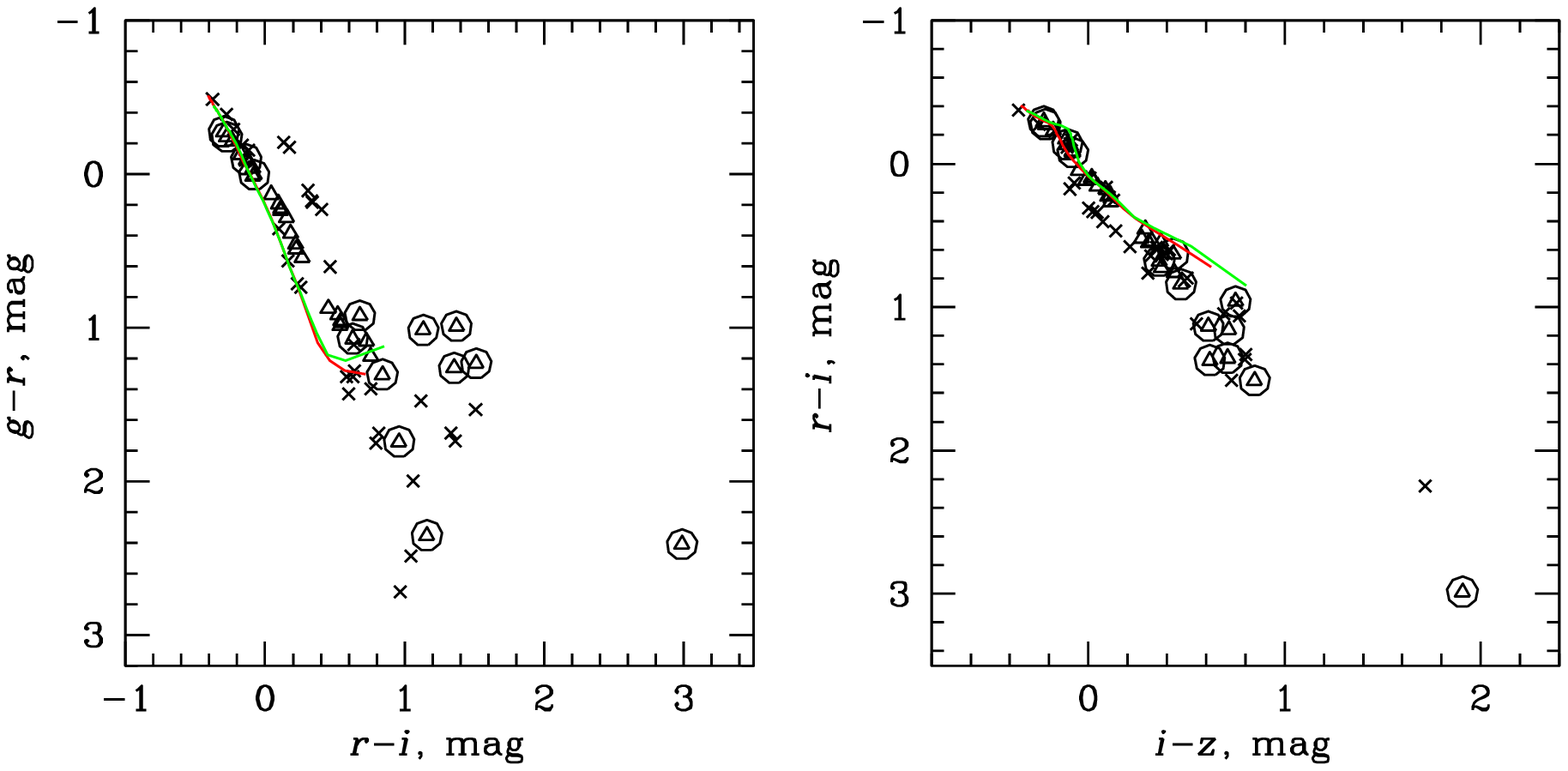}\\
\caption{Synthetic SDSS color-color diagrams derived from the MUSE 
(open circles) and XSL (open triangles) spectra. Larger open 
circles mark known variables, according to the SIMBAD database. and
although many statrs are variable, some distinct outliers are not.
Sequences for Solar abundance dwarfs (red line) and giant (green 
line) stars from \citet{1998ApJS..119..121L} are also shown. 
The extreme red outliers are IRAS\,15060+0947 (V*\,FV\,Boo) -- a 
known Mira variable. There are two points for this object on 
the right panel -- they correspond to the XSL and the MUSE 
spectra.}\label{fig:synth_colors}
\end{figure*}

The question remains, however, if the MUSE spectra have a more
reliable shape than the XSL spectra, because strictly speaking 
so far we have only demonstrated the good internal agreement 
between the six (or 12) individual MUSE observations. To provide
and external check we followed \citet{2014A&A...565A.117C}, and
calculated synthetic SDSS colors from both ours and the XSL 
spectra (Fig.\,\ref{fig:synth_colors}) using the {\it pyphot} 
tool\footnote{\url{http://mfouesneau.github.io/docs/pyphot/}}. 
The XSL spectra were median smoothed to remove outliers, e.g. due 
to poorly removed cosmic ray hits. The MUSE sequences are slightly 
tighter than the XSL ones, confirming that the MUSE spectra have 
more reliable shapes. This is expected, because of the slit losses 
and the imperfect order stitching of the XSL spectra. Furthermore,
X-shooter has three arms -- in effect, three different instruments, 
and some of the of the colors mix fluxes from different arms, which
may contribute to the larger scatter. A better spectral shape 
verification will be possible in the future with the {\it Gaia} 
low-resolution spectra.

\begin{figure}[!htb]
\centering
\includegraphics[width=9.0cm]{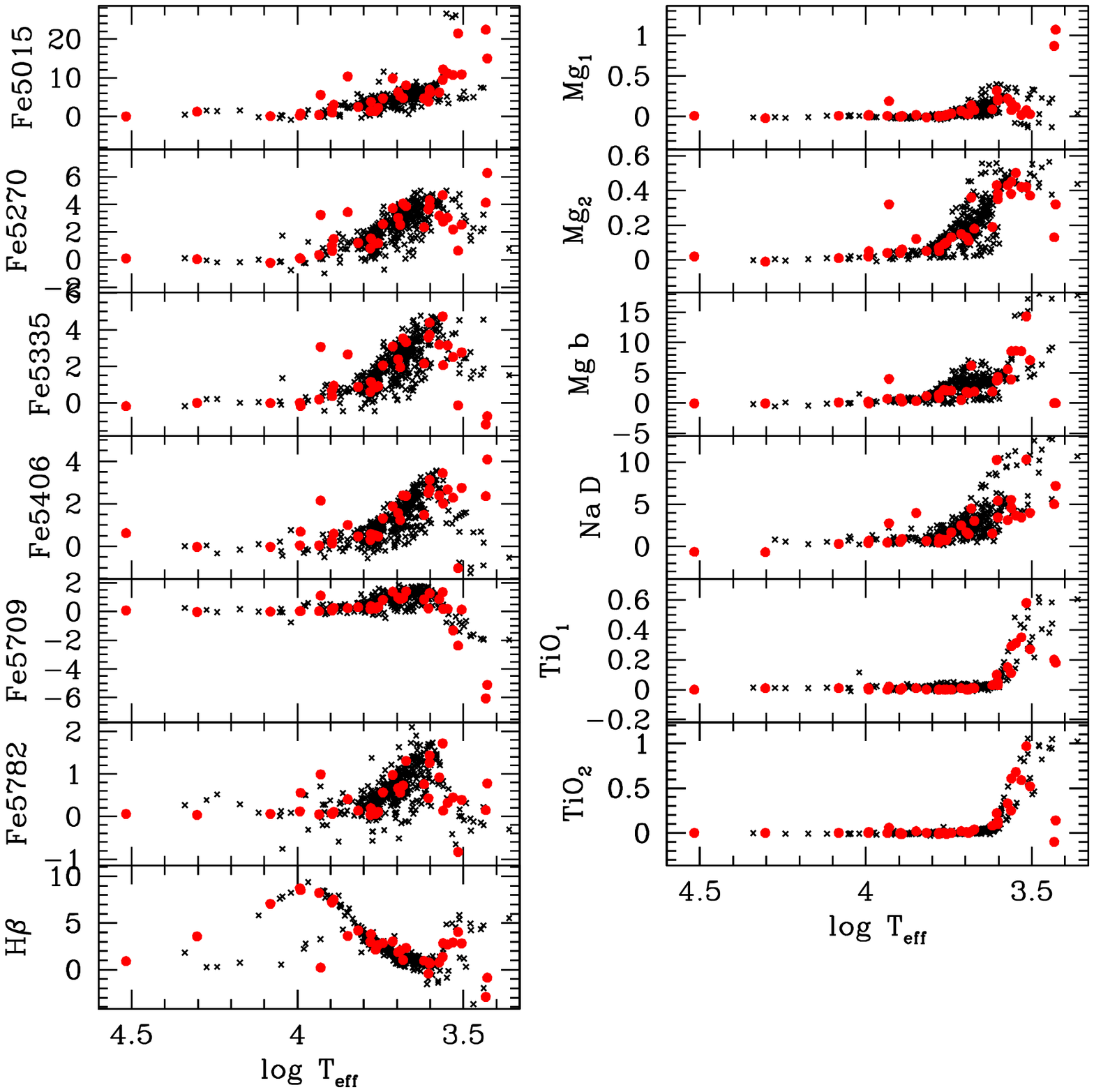}\\
\caption{Lick indices for the stars in our sample (red dots) and in 
the sample of \citet[][black dots]{1994ApJS...94..687W}. Following 
their definitions, Mg$_{\rm 1}$, Mg$_{\rm 2}$, TiO$_{\rm 1}$ and 
TiO$_{\rm 2}$ are in magnitudes, and the rest are equivalent widths 
in units of \AA. The two coldest objects that often deviate from the 
M-star dominated sequences are the carbon stars HD\,067507 and 
HD\,085405.}\label{fig:indices}
\end{figure}

The Lick indices \citep{1994ApJS...94..687W} that fall within the 
wavelength range covered by MUSE were measured in the new spectra 
(Table\,\ref{tab:indices}). This included: Fe5015, Fe5270, Fe5335, 
Fe5406, Fe5709, Fe5782, H$\beta$, Mg$_{\rm 1}$, Mg$_{\rm 2}$, Mg\,b, 
Na\,D, TiO$_{\rm 1}$ and TiO$_{\rm 2}$. As designed by our target 
selection, the measured values occupy the same locus as the Lick 
library (Fig.\,\ref{fig:indices}).

In the course of the analysis we noticed that the Lick indices of 
HD\,193281B correspond to a latter type than the F5:V: reported in 
Simbad. We derived a new spectral type of K2III using as templates 
our spectra of HD\,170820 and HD\,099998 and we adopted for this 
star the average of their effective temperatures, 
$T_{\rm eff}$=4354\,K with a tentative uncertainty of 57\,K -- the
larger of the uncertainties of the $T_{\rm eff}$ for these two 
stars.

The metal features of HD\,179821 are stronger than for other stars 
with similar temperature, but this is probably due to the supersolar
abundance of this star \citep{2016A&A...591A.118S}. Some Lick 
indices of late-M and C/S stars also deviate from the locus, but the 
spectra of these stars are dominated by broad molecular features, 
making the atomic indices such as Fe, Mg and H, meaningless.

\section{Summary and Conclusions}\label{sec:summary}

We present high signal-to-noise (S/N$>$70--200) MUSE spectra of 
35 stars across the spectral type sequence. The comparison with 
higher resolution existing data and spectral index measurements 
show reasonably good agreement, except for differences in the 
continuum shape that point at the real difficulties obtaining 
high-resolution spectra with wide spectral coverage: the 
instruments that deliver such kind of data spread the light over 
many orders and their combination is not trivial. Importantly, 
the integral field unit that we use does not suffer from slit 
losses.

The sample of spectra presented here is relatively limited in 
terms of number of stars, and to make this library more useful 
we need to populate more densely the parametric space. In 
particular, the metallicity range needs to be expanded. Our data 
suffer from the high blue wavelength limit of MUSE, missing some 
important CN, Ca and Fe spectral features in the 4100--4800\,\AA\ 
range. This is a hardware limitation that can only be addressed 
with other/future instruments. Further accurate broad band 
photometry is needed to extent the external verification of the 
continuum shape -- so far {\it Gaia}, SDSS and other photometric 
surveys provide measurements only for about a quarter of our 
sample stars -- mostly because our program stars are too bright. 
Expanding the MUSE library towards fainter stars will increase 
this fraction and make such a test statistically significant.

Despite these issues, our MUSE spectral library can be a useful 
tool for both stellar and galaxy research. This project started 
as a simple effort to complement the SXL DR1 library, but our 
spectra can be applied for various MUSE-based research -- they 
have an extra advantage of being obtained with the same instrument 
so the data format is the same, and any low-level instrumental 
signatures that might have remained in the data could cancel out. 
We plan to expand the number of library stars in the future.

\begin{acknowledgements}
This paper is based on observations made with the ESO Very Large 
Telescope at the La Silla Paranal Observatory under program
099.D-0623.
We have made extensive use of the SIMBAD Database at CDS (Centre 
de Donn\'ees astronomiques) Strasbourg and of the VizieR catalog 
access tool, CDS, Strasbourg, France. 
E.M.C., E.D.B., L.M., and A.P. acknowledge financial support from 
Padua University through grants DOR1715817/17, DOR1885254/18, 
DOR1935272/19, and BIRD164402/16.
We thank the referee for the comments that helped to improve the 
paper.
\end{acknowledgements}

% for the bibliography, at the end
\bibliographystyle{aa}
\bibliography{muse_std_104s}

\newpage
\appendix

\section{MUSE spectra.}

\begin{figure*}[]
\centering
\includegraphics[height=7.3cm]{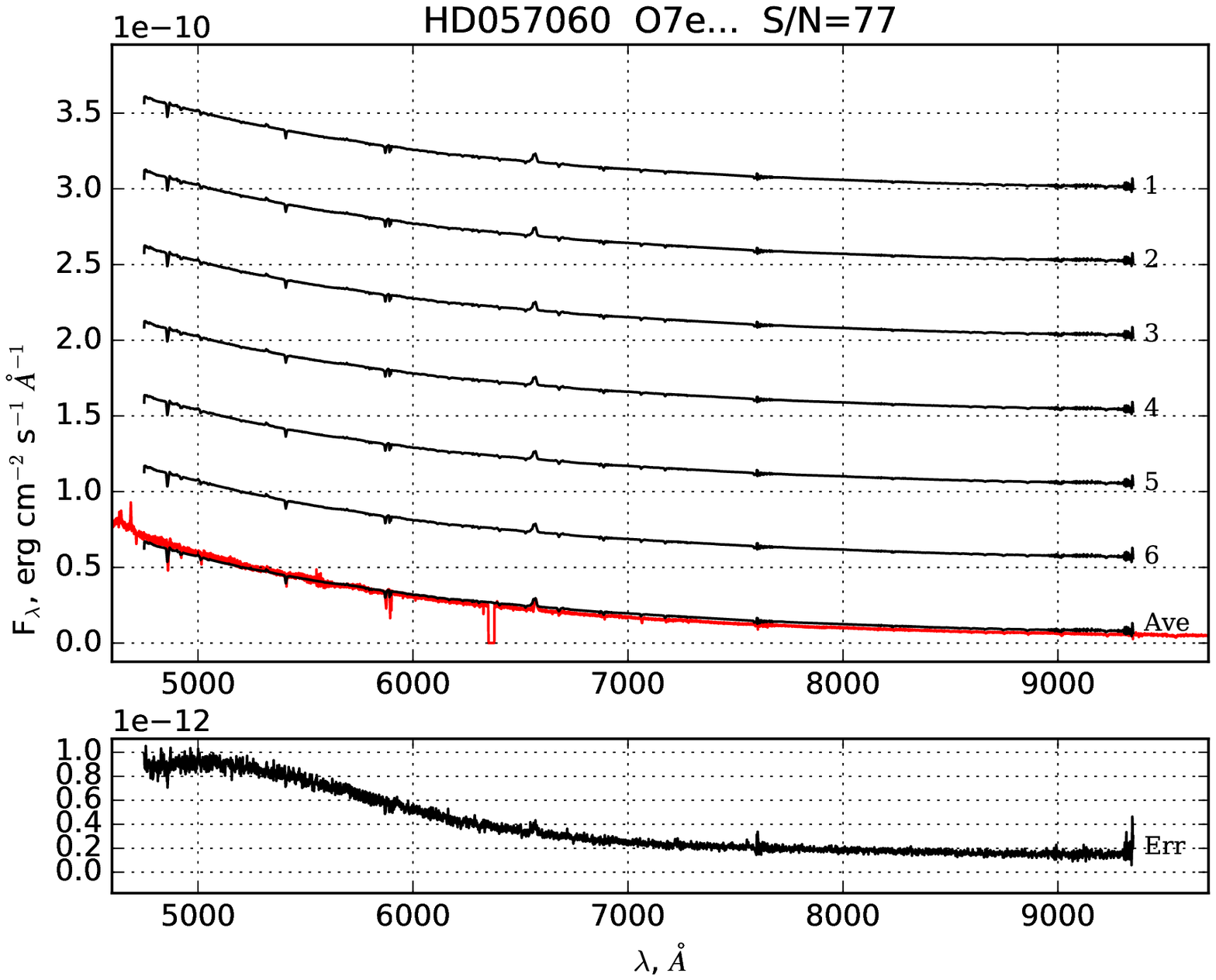} \includegraphics[height=7.3cm]{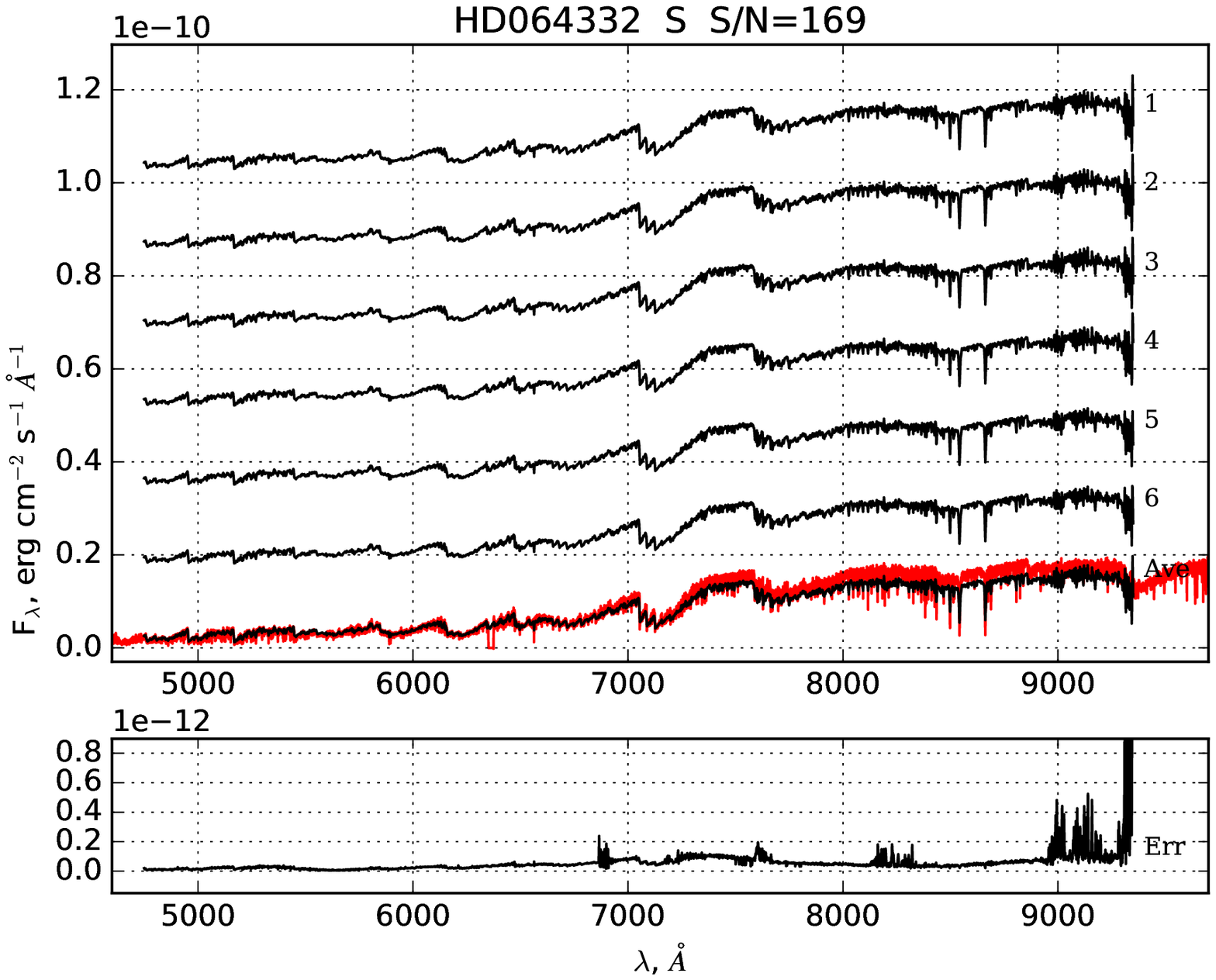} \\
\caption{MUSE spectra (labelled on the top of each panel). 
The upper sub-panels show the spectra extracted from each 
individual exposure (shifted up for clarity) and the average 
spectra of the object at its true flux level. The XSL 
spectra, when available, are plotted with red underneath 
the averaged MUSE spectrum.
The bottom sub-panels show the standard deviation of the 
average spectrum.
The spectra of the other stars are presented in the electronic 
edition only.}\label{fig:spectra_full_sample}
% \addtocounter{figure}{-1}
\end{figure*}

Table\,\ref{tab:obs_log} presents the log of our MUSE observations 
and Fig.\,\ref{fig:spectra_full_sample} -- the MUSE spectra.

\begin{table*}
\caption{Observing log. Six exposures were taken for all target 
except for HD\,204155 which was observed 12 times. The UT date 
and time at the start of the first exposure is listed, together 
with the airmass range for the entire sequence and the exposure 
time of each individual spectrum.}\label{tab:obs_log}
\begin{center}
\begin{small}
\begin{tabular}{@{ }l@{ }l@{ }c@{}c@{}c@{}c@{}c@{}r@{ }}
% \begin{tabular}{@{ }l@{ }l@{ }}
\hline\hline
ID & Alternative & RA DEC  & UT start, yyyy-       & sec\,$z$ & ~~~Exp.~~ & ~~Specphot.~ & sec\,$z$ \\
   & ID          & (J2000) & ~~~~~mm-dd hh:mm~~~~~ & dex      & sec       & Std.         &  dex      \\
(1) & (2) & (3) & (4) & (5) & (6) & (7) & (8) \\
\hline
HD\,057060              & ...           & 07:18:40.38\,$-$24:33:31.3 & 2017-05-03 01:07 & 1.63--1.70 &  0.14 & LTT\,3218 & 1.01 \\
HD\,064332              & ...           & 07:53:05.27\,$-$11:37:29.4 & 2017-05-03 01:23 & 1.61--1.68 &  4.80 & LTT\,3218 & 1.01 \\
HD\,067507              & RU\,Pup       & 08:07:29.83\,$-$22:54:45.3 & 2017-05-03 02:00 & 1.68--1.76 &  7.68 & LTT\,3218 & 1.01 \\ % RU\,Pup
HD\,085405              & Y\,Hya        & 09:51:03.72\,$-$23:01:02.3 & 2017-05-03 02:17 & 1.21--1.23 &  1.94 & LTT\,3218 & 1.01 \\ % Y\,Hya
HD\,096446              & ...           & 11:06:05.82\,$-$59:56:59.6 & 2017-05-03 01:41 & 1.24--1.24 &  1.94 & LTT\,3218 & 1.01 \\
HD\,099648              & ...           & 11:27:56.24\,$+$02:51:22.6 & 2017-07-18 23:54 & 1.89--2.00 &  0.14 & GD\,153   & 1.58 \\
HD\,099998              & BS\,4432      & 11:30:18.89\,$-$03:00:12.6 & 2017-05-03 00:39 & 1.10--1.09 &  0.15 & LTT\,3218 & 1.01 \\ % BS\,4432
HD\,100733              & BS\,4463      & 11:35:13.28\,$-$47:22:21.3 & 2017-05-02 06:55 & 2.39--2.52 &  0.95 & GD\,108   & 1.06 \\ % BS\,4463
HD\,306799              & CD$-$60\,3636~& 11:36:34.84\,$-$61:36:35.2 & 2017-05-02 23:17 & 1.38--1.37 &  3.86 & GD\,108   & 1.06 \\ % CD-60\,3636
HD\,101712              & ...           & 11:41:49.41\,$-$63:24:52.4 & 2017-04-18 06:59 & 1.84--1.89 &  4.79 & EG\,274   & 1.06 \\
HD\,102212              & BS\,4517      & 11:45:51.56\,$+$06:31:45.7 & 2017-05-03 00:53 & 1.20--1.19 &  0.15 & LTT\,3218 & 1.01 \\ % BS\,4517
HD\,114960              & ...           & 13:13:57.57\,$+$01:27:23.2 & 2017-04-01 07:35 & 1.35--1.39 &  1.93 & GD\,108   & 1.23 \\
IRAS\,15060+0947~       & ...           & 15:08:25.77\,$+$09:36:18.2 & 2017-07-18 23:35 & 1.22--1.21 & 43.65 & GD\,153   & 1.58 \\
HD\,147550              & ...           & 16:22:38.90\,$-$02:04:47.5 & 2017-05-21 04:53 & 1.08--1.08 &  1.95 & LTT\,7987 & 1.03 \\
HD\,160365              & ...           & 17:38:57.85\,$+$13:19:45.3 & 2017-05-21 08:33 & 1.53--1.57 &  1.92 & LTT\,7987 & 1.03 \\
HD\,160346              & ...           & 17:39:16.92\,$+$03:33:18.9 & 2017-05-21 06:06 & 1.14--1.14 &  1.94 & LTT\,7987 & 1.03 \\
HD\,163810              & ...           & 17:58:38.45\,$-$13:05:49.6 & 2017-05-21 08:51 & 1.18--1.21 & 14.56 & LTT\,7987 & 1.03 \\
HD\,164257              & ...           & 18:00:07.32\,$+$06:33:14.1 & 2017-05-21 09:07 & 1.45--1.49 &  1.92 & LTT\,7987 & 1.03 \\
\lbrack B86\rbrack\,133 & NSV\,24166    & 18:03:45.47\,$-$30:03:00.7 & 2017-05-02 07:11 & 1.03--1.02 &~83.77 & GD\,108   & 1.06 \\
HD\,167278              & ...           & 18:14:33.65\,$+$00:10:32.9 & 2017-05-21 09:22 & 1.35--1.39 &  7.72 & LTT\,7987 & 1.03 \\
HD\,170820              & ...           & 18:32:13.11\,$-$19:07:26.3 & 2017-05-28 09:43 & 1.31--1.32 &  3.87 & LTT\,7987 & 1.08 \\
HD\,172230              & ...           & 18:38:54.95\,$+$06:16:14.8 & 2017-05-31 05:09 & 1.28--1.26 &  3.87 & GD\,153   & 1.48 \\
HD\,173158              & ...           & 18:43:45.31\,$+$05:44:14.6 & 2017-05-31 05:25 & 1.25--1.23 &  6.78 & GD\,153   & 1.48 \\
HD\,174966              & ...           & 18:53:07.83\,$+$01:45:19.7 & 2017-05-31 05:41 & 1.19--1.17 &  4.85 & GD\,153   & 1.48 \\
HD\,175640              & ...           & 18:56:22.66\,$-$01:47:59.5 & 2017-05-21 09:36 & 1.23--1.26 &  1.94 & LTT\,7987 & 1.03 \\
HD\,179821              & ...           & 19:13:58.61\,$+$00:07:31.9 & 2017-05-31 05:56 & 1.18--1.17 &  7.76 & GD\,153   & 1.48 \\
HD\,232078              & ...           & 19:38:12.07\,$+$16:48:25.6 & 2017-05-31 07:12 & 1.35--1.34 &  9.67 & GD\,153   & 1.48 \\
HD\,193256              & ...           & 20:20:26.57\,$-$29:11:28.8 & 2017-05-31 06:10 & 1.16--1.14 &  1.95 & GD\,153   & 1.48 \\
HD\,193281A             & ...           & 20:20:27.88\,$-$29:11:50.0 & 2017-05-31 06:10 & 1.16--1.14 &  1.95 & GD\,153   & 1.48 \\
HD\,193281B             & ...           & 20:20:28.07\,$-$29:11:47.2 & 2017-05-31 06:10 & 1.16--1.14 &  1.95 & GD\,153   & 1.48 \\
HD\,193896              & ...           & 20:23:00.79\,$-$09:39:17.0 & 2017-05-31 06:25 & 1.19--1.17 &  1.94 & GD\,153   & 1.48 \\
HD\,196892              & ...           & 20:40:49.38\,$-$18:47:33.3 & 2017-05-31 06:42 & 1.15--1.13 &  7.78 & GD\,153   & 1.48 \\
HD\,200081              & ...           & 21:01:22.42\,$-$02:30:50.4 & 2017-05-31 06:55 & 1.28--1.25 &  6.77 & GD\,153   & 1.48 \\
HD\,204155              & ...           & 21:26:42.91\,$+$05:26:29.9 & 2017-05-31 07:26 & 1.37--1.29 &  7.72 & GD\,153   & 1.48 \\
HD\,209290              & ...           & 22:02:10.27\,$+$01:24:00.8 & 2017-05-31 07:56 & 1.33--1.30 &  9.66 & GD\,153   & 1.48 \\
\hline
\end{tabular}
\end{small}
\end{center}
\end{table*}

\begin{table}
\caption{MUSE spectra of the program stars. Only ten entries 
for a few spectra are shown for guidance. The full spectra are
available in the electronic edition.}\label{table:full_spectra}
\begin{center}
\begin{small}
\begin{tabular}{@{ }l@{ }c@{ }c@{ }}
\hline\hline
$\lambda$ & F$_\lambda$, $\sigma$(F$_\lambda$) \\
$\AA$     & erg\,cm$^{-1}$\,s$^{-1}$\,$\AA^{-1}$ \\
(1)       & (2) \\
\hline
\multicolumn{2}{c}{\lbrack B86\rbrack\,133} \\
4750.351 & 9.0321e-15 1.3444e-16 \\
4751.601 & 9.2171e-15 1.7869e-16 \\
4752.851 & 9.1116e-15 1.3160e-16 \\
4754.101 & 9.0764e-15 1.4030e-16 \\
4755.351 & 9.3078e-15 1.4568e-16 \\
% 4756.601 & 9.3031e-15 1.5083e-16 \\
% 4757.851 & 8.5689e-15 1.0591e-16 \\
% 4759.101 & 7.6297e-15 9.5282e-17 \\
% 4760.351 & 6.8822e-15 1.0720e-16 \\
% 4761.601 & 5.9430e-15 1.0564e-16 \\
\multicolumn{2}{c}{...} \\
\multicolumn{2}{c}{HD\,057060} \\
4749.690 & 6.2487e-11 9.8514e-13 \\
4750.940 & 6.6745e-11 9.7720e-13 \\
4752.190 & 6.7090e-11 8.6972e-13 \\
4753.440 & 6.6901e-11 9.2688e-13 \\
4754.690 & 6.7012e-11 9.6207e-13 \\
% 4755.940 & 6.7031e-11 8.5821e-13 \\
% 4757.190 & 6.6923e-11 1.0520e-12 \\
% 4758.440 & 6.6567e-11 9.7377e-13 \\
% 4759.690 & 6.6468e-11 9.0956e-13 \\
% 4760.940 & 6.6522e-11 8.7795e-13 \\
\multicolumn{2}{c}{...} \\
% \multicolumn{2}{c}{HD\,064332} \\
% 4749.609 & 2.5791e-12 2.0012e-14 \\
% 4750.859 & 2.6738e-12 1.4487e-14 \\
% 4752.109 & 2.6184e-12 1.6491e-14 \\
% 4753.359 & 2.6216e-12 1.2228e-14 \\
% 4754.609 & 2.7259e-12 1.6599e-14 \\
% 4755.859 & 2.7450e-12 1.8022e-14 \\
% 4757.109 & 2.5564e-12 1.4855e-14 \\
% 4758.359 & 2.2996e-12 9.9624e-15 \\
% 4759.609 & 2.1371e-12 1.0996e-14 \\
% 4760.859 & 1.8722e-12 7.0553e-15 \\
% \multicolumn{2}{c}{...} \\
\hline
\end{tabular}
\end{small}
\end{center}
\end{table}

\newpage
\section{Comparison with the XSL spectra.}

Table\,\ref{tab:ratio_fits} lists the coefficients of polynomial 
fits to the ratios of the XSL spectra to our MUSE spectra. For 
further details see Sec.\,\ref{sec:analysis}.

\begin{table*}
\caption{Coefficients and their errors of second order polynomial 
fits to the ratios of the XSL spectra to our MUSE spectra: 
$Ratio = a_0 + a_1 \times \lambda + a_2 \times \lambda^2$.
The standard deviation of the residuals $\sigma$ is also 
listed.}\label{tab:ratio_fits}
\begin{center}
\begin{small}
\begin{tabular}{@{ }l@{ }c@{ }c@{ }c@{ }c@{ }}
\hline\hline
ID & $a_0$ & $a_1$ & $a_2$ & $\sigma$ \\
(1) & (2) & (3) & (4) & (5) \\
\hline
      [B86]\,133 &  1.0875e+01 $\pm$ 4.9106e-01 & -2.6034e-03 $\pm$ 1.4250e-04 &  2.1119e-07 $\pm$ 1.0070e-08 & 8.9657e-01 \\
      HD\,057060 &  7.7783e+00 $\pm$ 1.5925e-01 &  4.1637e-04 $\pm$ 4.6240e-05 &  5.4824e-08 $\pm$ 3.2693e-09 & 2.9142e-01 \\
      HD\,064332 &  3.5697e+01 $\pm$ 3.4877e-01 & -3.9873e-03 $\pm$ 1.0127e-04 &  2.0084e-07 $\pm$ 7.1597e-09 & 6.3832e-01 \\
      HD\,067507 & -2.1863e+00 $\pm$ 6.3861e-01 &  3.2643e-03 $\pm$ 1.8542e-04 & -1.9423e-07 $\pm$ 1.3110e-08 & 1.1688e+00 \\
      HD\,085405 & -2.1686e+01 $\pm$ 3.6461e-01 &  7.1684e-03 $\pm$ 1.0587e-04 & -4.3791e-07 $\pm$ 7.4850e-09 & 6.6732e-01 \\
      HD\,096446 &  9.9209e-01 $\pm$ 7.7207e-03 & -1.4847e-05 $\pm$ 2.2418e-06 &  1.3103e-09 $\pm$ 1.5850e-10 & 1.4131e-02 \\
      HD\,099648 &  5.1713e+00 $\pm$ 6.3635e-02 &  3.6673e-04 $\pm$ 1.8477e-05 & -2.7750e-08 $\pm$ 1.3064e-09 & 1.1647e-01 \\
      HD\,099998 &  6.1662e+00 $\pm$ 7.5652e-02 &  3.3898e-05 $\pm$ 2.1966e-05 & -1.2490e-08 $\pm$ 1.5530e-09 & 1.3847e-01 \\
      HD\,100733 &  5.1682e+01 $\pm$ 4.3041e-01 & -8.1477e-03 $\pm$ 1.2497e-04 &  6.4140e-07 $\pm$ 8.8351e-09 & 7.8776e-01 \\
      HD\,101712 &  1.6022e+01 $\pm$ 2.3634e-01 & -7.1526e-04 $\pm$ 6.8618e-05 &  4.8079e-08 $\pm$ 4.8512e-09 & 4.3258e-01 \\
      HD\,101712 &  6.7516e+00 $\pm$ 2.2716e-01 &  1.4180e-03 $\pm$ 6.5952e-05 & -8.2102e-08 $\pm$ 4.6627e-09 & 4.1578e-01 \\
      HD\,102212 &  5.0144e+01 $\pm$ 8.6927e-01 & -3.0969e-03 $\pm$ 2.5240e-04 &  3.6520e-07 $\pm$ 1.7845e-08 & 1.5910e+00 \\
      HD\,114960 &  4.0233e+01 $\pm$ 9.7355e-01 & -5.0542e-03 $\pm$ 2.5999e-04 &  4.4436e-07 $\pm$ 1.7128e-08 & 7.8302e-01 \\
      HD\,147550 &  2.1748e+00 $\pm$ 2.1553e-02 &  2.0414e-04 $\pm$ 6.2580e-06 & -1.4080e-08 $\pm$ 4.4245e-10 & 3.9450e-02 \\
      HD\,160346 &  5.7738e+00 $\pm$ 6.3278e-02 & -1.7415e-04 $\pm$ 1.8292e-05 &  1.1002e-08 $\pm$ 1.2900e-09 & 1.1465e-01 \\
      HD\,160365 &  4.5644e+00 $\pm$ 3.9193e-02 &  3.6423e-05 $\pm$ 1.1329e-05 & -4.6190e-09 $\pm$ 7.9902e-10 & 7.1011e-02 \\
      HD\,163810 &  3.6215e+00 $\pm$ 4.3395e-02 & -7.4719e-05 $\pm$ 1.2544e-05 &  2.6363e-09 $\pm$ 8.8470e-10 & 7.8619e-02 \\
      HD\,164257 &  2.9161e+00 $\pm$ 2.1773e-02 & -1.9122e-04 $\pm$ 6.3208e-06 &  1.3446e-08 $\pm$ 4.4681e-10 & 3.9867e-02 \\
      HD\,167278 &  7.6624e+00 $\pm$ 5.1776e-02 & -4.5347e-04 $\pm$ 1.4967e-05 &  2.0024e-08 $\pm$ 1.0555e-09 & 9.3807e-02 \\
      HD\,170820 &  5.0225e+00 $\pm$ 3.4454e-02 & -6.4631e-04 $\pm$ 1.0002e-05 &  4.1002e-08 $\pm$ 7.0704e-10 & 6.3086e-02 \\
      HD\,172230 &  2.4422e+00 $\pm$ 2.7921e-02 &  1.0172e-04 $\pm$ 8.0709e-06 & -5.0514e-09 $\pm$ 5.6920e-10 & 5.0586e-02 \\
      HD\,173158 &  8.3642e+00 $\pm$ 4.7810e-02 & -1.5581e-03 $\pm$ 1.3820e-05 &  1.0331e-07 $\pm$ 9.7467e-10 & 8.6622e-02 \\
      HD\,174966 &  4.5597e+00 $\pm$ 1.9866e-02 & -3.4119e-04 $\pm$ 5.7425e-06 &  2.3893e-08 $\pm$ 4.0499e-10 & 3.5995e-02 \\
      HD\,175640 &  2.5360e+00 $\pm$ 1.8827e-02 &  3.5332e-05 $\pm$ 5.4655e-06 & -2.5504e-09 $\pm$ 3.8634e-10 & 3.4474e-02 \\
      HD\,179821 &  2.8459e+00 $\pm$ 9.0632e-02 & -1.1607e-04 $\pm$ 2.6310e-05 &  1.3874e-08 $\pm$ 1.8598e-09 & 1.6596e-01 \\
     HD\,193281A &  2.9758e+00 $\pm$ 1.7990e-02 & -3.8776e-05 $\pm$ 5.2226e-06 &  5.5744e-09 $\pm$ 3.6918e-10 & 3.2940e-02 \\
      HD\,193896 &  3.5954e+00 $\pm$ 2.2904e-02 & -2.7330e-04 $\pm$ 6.6490e-06 &  1.9823e-08 $\pm$ 4.7000e-10 & 4.1936e-02 \\
      HD\,196892 &  3.9280e+00 $\pm$ 2.3032e-02 & -4.0473e-04 $\pm$ 6.6862e-06 &  2.8704e-08 $\pm$ 4.7263e-10 & 4.2171e-02 \\
      HD\,200081 &  2.0961e+00 $\pm$ 2.5504e-02 &  1.2569e-04 $\pm$ 7.4035e-06 & -5.3440e-09 $\pm$ 5.2332e-10 & 4.6702e-02 \\
      HD\,204155 &  3.0281e+00 $\pm$ 2.9675e-02 & -8.6000e-05 $\pm$ 8.5776e-06 &  4.6207e-09 $\pm$ 6.0491e-10 & 5.3764e-02 \\
      HD\,209290 & -3.2922e+00 $\pm$ 5.5030e-02 &  1.6589e-03 $\pm$ 1.5975e-05 & -9.6328e-08 $\pm$ 1.1292e-09 & 1.0076e-01 \\
      HD\,232078 &  2.5877e+00 $\pm$ 8.0086e-02 & -5.7710e-05 $\pm$ 2.3251e-05 &  9.4294e-09 $\pm$ 1.6436e-09 & 1.4658e-01 \\
      HD\,306799 &  1.5678e+01 $\pm$ 2.1450e-01 & -1.0232e-03 $\pm$ 6.2275e-05 &  5.6597e-08 $\pm$ 4.4024e-09 & 3.9253e-01 \\
IRAS\,15060+0947 & -1.0062e+01 $\pm$ 7.1662e+00 & -4.3525e-03 $\pm$ 1.9079e-03 &  1.4548e-06 $\pm$ 1.2536e-07 & 5.5817e+00 \\
\hline
\end{tabular}
\end{small}
\end{center}
\end{table*}

% \newpage
\section{Lick indices from the MUSE spectra.}

Table\,\ref{tab:indices} shows the Lick indices measured on our 
MUSE spectra.

\newpage
\begin{sidewaystable}
% \begin{table*}
\caption{Lick indices measured on the MUSE spectra. 
Following \citet{1994ApJS...94..687W}, Mg$_{\rm 1}$, Mg$_{\rm 2}$, 
TiO$_{\rm 1}$ and TiO$_{\rm 2}$ are in magnitudes, and the rest and 
the rest are equivalent widths in units of \AA.}\label{tab:indices}
% \begin{center}
\begin{small}
% \begin{tabular}{@{}l@{ }c@{ }c@{ }c@{ }c@{ }c@{ }c@{ }c@{ }c@{ }c@{ }c@{ }c@{ }c@{ }c@{}}
\begin{tabular}{@{}l@{ }r@{ }r@{ }r@{ }r@{ }r@{ }r@{ }r@{ }r@{ }r@{ }r@{ }r@{ }r@{ }r@{}}
\hline\hline
ID & \multicolumn{1}{c}{Fe5015} & \multicolumn{1}{c}{Fe5270} & \multicolumn{1}{c}{Fe5335} & \multicolumn{1}{c}{Fe5406} & \multicolumn{1}{c}{Fe5709} & \multicolumn{1}{c}{Fe5782} & \multicolumn{1}{c}{H$\beta$} & \multicolumn{1}{c}{Mg$_{\rm 1}$} & \multicolumn{1}{c}{Mg$_{\rm 2}$} & \multicolumn{1}{c}{Mg\,b} & \multicolumn{1}{c}{Na\,D} & \multicolumn{1}{c}{TiO$_{\rm 1}$} & \multicolumn{1}{c}{TiO$_{\rm 2}$} \\
\hline
      [B86]\,133 &  12.14$\pm$0.19  &   2.76$\pm$0.11  &   2.07$\pm$0.11  &   2.02$\pm$0.09  &   0.18$\pm$0.06  &   0.14$\pm$0.05  &   2.83$\pm$0.12  &   0.081$\pm$0.002  &   0.447$\pm$0.003  &   8.57$\pm$0.08  &   4.57$\pm$0.06  & ~~0.289$\pm$0.001  &   0.608$\pm$0.001 \\
        HD057060 &   0.05$\pm$0.30  &   0.11$\pm$0.17  &$-$0.17$\pm$0.19  &   0.62$\pm$0.16  &   0.08$\pm$0.13  &   0.06$\pm$0.12  &   0.90$\pm$0.12  &   0.012$\pm$0.003  &   0.015$\pm$0.004  &$-$0.07$\pm$0.15  &$-$0.64$\pm$0.17  &   0.003$\pm$0.003  &   0.000$\pm$0.003 \\
        HD064332 &  10.77$\pm$0.09  &   2.19$\pm$0.08  &   2.51$\pm$0.08  &   2.29$\pm$0.05  &$-$1.32$\pm$0.02  &   0.45$\pm$0.02  &   2.92$\pm$0.04  &   0.024$\pm$0.001  &   0.422$\pm$0.002  &   8.57$\pm$0.05  &   3.43$\pm$0.04  &   0.354$\pm$0.001  &   0.593$\pm$0.002 \\
        HD067507 &  14.99$\pm$0.14  &   6.26$\pm$0.08  &$-$0.71$\pm$0.11  &   4.08$\pm$0.07  &$-$5.11$\pm$0.08  &   0.78$\pm$0.06  &$-$0.85$\pm$0.09  &   1.074$\pm$0.002  &   0.317$\pm$0.001  &   0.00$\pm$0.13  &   7.18$\pm$0.08  &   0.181$\pm$0.002  &   0.135$\pm$0.002 \\
        HD085405 &  22.38$\pm$0.11  &   4.13$\pm$0.05  &$-$1.16$\pm$0.06  &   2.36$\pm$0.05  &$-$6.05$\pm$0.05  &   0.15$\pm$0.03  &$-$2.91$\pm$0.10  &   0.867$\pm$0.001  &   0.131$\pm$0.001  &   0.00$\pm$0.10  &   5.02$\pm$0.04  &   0.203$\pm$0.001  &$-$0.100$\pm$0.001 \\
        HD096446 &   1.27$\pm$0.16  &   0.05$\pm$0.13  &   0.01$\pm$0.15  &$-$0.03$\pm$0.12  &$-$0.03$\pm$0.06  &   0.04$\pm$0.05  &   3.56$\pm$0.05  &$-$0.015$\pm$0.002  &$-$0.009$\pm$0.003  &$-$0.04$\pm$0.12  &$-$0.68$\pm$0.09  &   0.007$\pm$0.002  &$-$0.001$\pm$0.002 \\
        HD099648 &   6.27$\pm$0.13  &   3.03$\pm$0.05  &   2.38$\pm$0.05  &   1.56$\pm$0.05  &   0.99$\pm$0.05  &   0.67$\pm$0.05  &   1.88$\pm$0.06  &   0.044$\pm$0.001  &   0.128$\pm$0.001  &   1.83$\pm$0.05  &   1.68$\pm$0.06  &   0.005$\pm$0.001  &   0.013$\pm$0.001 \\
        HD099998 &   6.02$\pm$0.16  &   3.89$\pm$0.08  &   3.73$\pm$0.09  &   2.68$\pm$0.07  &   1.27$\pm$0.06  &   1.26$\pm$0.06  &   0.65$\pm$0.09  &   0.197$\pm$0.002  &   0.347$\pm$0.002  &   3.81$\pm$0.08  &   3.44$\pm$0.09  &   0.047$\pm$0.002  &   0.130$\pm$0.002 \\
        HD100733 &  11.12$\pm$0.19  &   3.04$\pm$0.09  &   3.14$\pm$0.08  &   2.68$\pm$0.07  &   0.19$\pm$0.07  &   0.32$\pm$0.07  &   2.67$\pm$0.10  &   0.121$\pm$0.002  &   0.499$\pm$0.003  &   8.65$\pm$0.08  &   3.69$\pm$0.09  &   0.312$\pm$0.002  &   0.675$\pm$0.003 \\
        HD101712 &  10.86$\pm$0.35  &   2.53$\pm$0.24  &   2.76$\pm$0.25  &   2.75$\pm$0.21  &   0.13$\pm$0.20  &   0.39$\pm$0.19  &   2.81$\pm$0.20  &   0.029$\pm$0.004  &   0.374$\pm$0.006  &   7.11$\pm$0.16  &   3.99$\pm$0.26  &   0.267$\pm$0.006  &   0.523$\pm$0.007 \\
        HD102212 &   6.24$\pm$0.39  &   3.19$\pm$0.23  &   3.19$\pm$0.24  &   2.40$\pm$0.19  &   0.86$\pm$0.13  &   0.92$\pm$0.12  &   0.77$\pm$0.17  &   0.220$\pm$0.004  &   0.431$\pm$0.006  &   5.59$\pm$0.20  &   3.11$\pm$0.18  &   0.147$\pm$0.004  &   0.331$\pm$0.005 \\
        HD114960 &   6.93$\pm$0.16  &   4.31$\pm$0.10  &   4.38$\pm$0.10  &   3.12$\pm$0.08  &   1.28$\pm$0.05  &   1.44$\pm$0.03  &   0.74$\pm$0.04  &   0.221$\pm$0.002  &   0.380$\pm$0.003  &   4.31$\pm$0.10  &   5.43$\pm$0.04  &   0.035$\pm$0.001  &   0.104$\pm$0.002 \\
        HD147550 &   0.24$\pm$0.20  &   0.13$\pm$0.02  &   0.01$\pm$0.01  &   0.05$\pm$0.02  &   0.02$\pm$0.05  &   0.12$\pm$0.05  &   8.72$\pm$0.08  &   0.009$\pm$0.001  &   0.021$\pm$0.002  &   0.26$\pm$0.05  &   0.42$\pm$0.09  &   0.008$\pm$0.002  &   0.004$\pm$0.002 \\
        HD160346 &   4.80$\pm$0.15  &   4.06$\pm$0.09  &   3.53$\pm$0.09  &   2.40$\pm$0.07  &   0.95$\pm$0.01  &   0.73$\pm$0.03  &   1.04$\pm$0.08  &   0.138$\pm$0.002  &   0.360$\pm$0.002  &   6.17$\pm$0.08  &   4.52$\pm$0.06  &   0.010$\pm$0.002  &   0.017$\pm$0.002 \\
        HD160365 &   3.75$\pm$0.09  &   1.54$\pm$0.12  &   1.16$\pm$0.14  &   0.59$\pm$0.11  &   0.37$\pm$0.04  &   0.20$\pm$0.06  &   3.80$\pm$0.06  &   0.001$\pm$0.002  &   0.049$\pm$0.002  &   0.84$\pm$0.11  &   0.92$\pm$0.11  &   0.003$\pm$0.003  &   0.000$\pm$0.002 \\
        HD163810 &   1.46$\pm$0.15  &   1.16$\pm$0.07  &   0.84$\pm$0.07  &   0.51$\pm$0.06  &   0.20$\pm$0.06  &   0.06$\pm$0.06  &   2.14$\pm$0.07  &   0.011$\pm$0.001  &   0.086$\pm$0.002  &   2.20$\pm$0.06  &   0.77$\pm$0.09  &   0.003$\pm$0.002  &$-$0.000$\pm$0.002 \\
        HD164257 &   0.84$\pm$0.24  &   0.08$\pm$0.04  &$-$0.15$\pm$0.07  &   0.69$\pm$0.07  &   0.03$\pm$0.09  &   0.56$\pm$0.09  &   8.53$\pm$0.05  &   0.017$\pm$0.002  &   0.048$\pm$0.002  &$-$0.07$\pm$0.05  &   0.68$\pm$0.12  &   0.001$\pm$0.003  &   0.009$\pm$0.001 \\
        HD167278 &   2.48$\pm$0.11  &   1.22$\pm$0.08  &   0.89$\pm$0.10  &   0.47$\pm$0.10  &   0.31$\pm$0.06  &   0.14$\pm$0.04  &   4.21$\pm$0.03  &$-$0.006$\pm$0.001  &   0.045$\pm$0.001  &   1.16$\pm$0.05  &   0.64$\pm$0.03  &   0.004$\pm$0.001  &$-$0.001$\pm$0.001 \\
        HD170820 &   8.07$\pm$0.17  &   3.87$\pm$0.14  &   3.32$\pm$0.15  &   2.37$\pm$0.14  &   1.40$\pm$0.13  &   1.31$\pm$0.13  &   2.32$\pm$0.06  &   0.083$\pm$0.002  &   0.185$\pm$0.003  &   1.90$\pm$0.13  &   3.04$\pm$0.17  &   0.011$\pm$0.004  &   0.042$\pm$0.002 \\
        HD172230 &   3.10$\pm$0.07  &   1.52$\pm$0.06  &   0.95$\pm$0.07  &   0.60$\pm$0.06  &   0.25$\pm$0.05  &   0.11$\pm$0.04  &   7.50$\pm$0.03  &   0.015$\pm$0.001  &   0.056$\pm$0.001  &   0.25$\pm$0.06  &   0.88$\pm$0.04  &   0.006$\pm$0.001  &$-$0.005$\pm$0.001 \\
        HD173158 &   9.86$\pm$0.07  &   3.73$\pm$0.05  &   3.09$\pm$0.08  &   1.88$\pm$0.09  &   1.38$\pm$0.06  &   0.98$\pm$0.04  &   3.04$\pm$0.03  &   0.069$\pm$0.001  &   0.146$\pm$0.001  &   0.52$\pm$0.02  &   2.50$\pm$0.04  &   0.009$\pm$0.001  &   0.024$\pm$0.001 \\
        HD174966 &   1.85$\pm$0.18  &   1.02$\pm$0.10  &   0.67$\pm$0.09  &   0.27$\pm$0.07  &   0.14$\pm$0.03  &   0.08$\pm$0.04  &   7.21$\pm$0.05  &$-$0.003$\pm$0.002  &   0.043$\pm$0.002  &   0.68$\pm$0.10  &   0.64$\pm$0.07  &   0.004$\pm$0.002  &$-$0.007$\pm$0.002 \\
        HD175640 &   0.14$\pm$0.17  &$-$0.22$\pm$0.11  &$-$0.00$\pm$0.11  &$-$0.02$\pm$0.08  &$-$0.01$\pm$0.03  &   0.06$\pm$0.04  &   7.05$\pm$0.05  &   0.008$\pm$0.002  &   0.010$\pm$0.002  &   0.11$\pm$0.10  &   0.28$\pm$0.07  &   0.006$\pm$0.002  &$-$0.001$\pm$0.001 \\
        HD179821 &  10.34$\pm$0.11  &   3.44$\pm$0.06  &   2.66$\pm$0.05  &   1.01$\pm$0.03  &   0.24$\pm$0.03  &   0.41$\pm$0.03  &   3.61$\pm$0.04  &   0.023$\pm$0.001  &   0.123$\pm$0.002  &   0.31$\pm$0.07  &   3.97$\pm$0.05  &   0.010$\pm$0.001  &   0.017$\pm$0.001 \\
        HD193256 &   0.99$\pm$0.16  &   0.64$\pm$0.08  &   0.38$\pm$0.09  &   0.14$\pm$0.08  &   0.07$\pm$0.07  &   0.06$\pm$0.06  &   7.28$\pm$0.05  &   0.004$\pm$0.002  &   0.045$\pm$0.002  &   0.76$\pm$0.07  &   0.53$\pm$0.09  &   0.004$\pm$0.002  &$-$0.004$\pm$0.002 \\
       HD193281A &   0.40$\pm$0.14  &   0.36$\pm$0.08  &   0.20$\pm$0.09  &   0.05$\pm$0.08  &   0.03$\pm$0.07  &   0.05$\pm$0.06  &   8.22$\pm$0.05  &   0.006$\pm$0.001  &   0.039$\pm$0.002  &   0.72$\pm$0.07  &   0.46$\pm$0.09  &   0.004$\pm$0.002  &$-$0.004$\pm$0.002 \\
       HD193281B &   5.55$\pm$2.43  &   3.24$\pm$1.13  &   3.06$\pm$1.18  &   2.15$\pm$0.97  &   1.11$\pm$0.74  &   0.99$\pm$0.67  &   0.21$\pm$1.13  &   0.189$\pm$0.026  &   0.325$\pm$0.032  &   3.98$\pm$1.13  &   2.74$\pm$0.91  &   0.019$\pm$0.020  &   0.057$\pm$0.017 \\
        HD193896 &   5.26$\pm$0.09  &   2.52$\pm$0.04  &   1.96$\pm$0.04  &   1.24$\pm$0.04  &   0.86$\pm$0.05  &   0.56$\pm$0.06  &   2.04$\pm$0.03  &   0.031$\pm$0.001  &   0.110$\pm$0.001  &   1.65$\pm$0.04  &   1.45$\pm$0.08  &   0.005$\pm$0.002  &   0.002$\pm$0.002 \\
        HD196892 &   1.40$\pm$0.16  &   0.83$\pm$0.07  &   0.61$\pm$0.08  &   0.30$\pm$0.06  &   0.17$\pm$0.03  &   0.04$\pm$0.02  &   2.98$\pm$0.08  &   0.014$\pm$0.002  &   0.068$\pm$0.002  &   1.41$\pm$0.08  &   0.48$\pm$0.02  &   0.002$\pm$0.000  &$-$0.007$\pm$0.001 \\
        HD200081 &   4.69$\pm$0.19  &   2.57$\pm$0.08  &   2.06$\pm$0.07  &   1.32$\pm$0.05  &   0.81$\pm$0.04  &   0.57$\pm$0.04  &   2.83$\pm$0.08  &   0.036$\pm$0.002  &   0.127$\pm$0.002  &   2.05$\pm$0.09  &   1.67$\pm$0.06  &   0.004$\pm$0.001  &   0.003$\pm$0.002 \\
        HD204155 &   2.26$\pm$0.20  &   1.21$\pm$0.13  &   0.92$\pm$0.15  &   0.47$\pm$0.13  &   0.32$\pm$0.10  &   0.11$\pm$0.08  &   2.65$\pm$0.10  &   0.019$\pm$0.002  &   0.096$\pm$0.003  &   2.12$\pm$0.11  &   0.76$\pm$0.10  &   0.003$\pm$0.002  &$-$0.006$\pm$0.002 \\
        HD209290 &   3.85$\pm$0.10  &   3.63$\pm$0.08  &   3.58$\pm$0.10  &   2.51$\pm$0.08  &   0.21$\pm$0.04  &   0.43$\pm$0.03  &$-$0.43$\pm$0.05  &   0.322$\pm$0.001  &   0.430$\pm$0.002  &   3.77$\pm$0.06  &  10.27$\pm$0.03  &   0.098$\pm$0.001  &   0.220$\pm$0.001 \\
        HD232078 &   4.70$\pm$0.10  &   2.36$\pm$0.06  &   2.15$\pm$0.07  &   1.48$\pm$0.05  &   0.84$\pm$0.03  &   0.76$\pm$0.02  &   0.95$\pm$0.04  &   0.091$\pm$0.001  &   0.192$\pm$0.001  &   1.90$\pm$0.03  &   1.53$\pm$0.03  &   0.031$\pm$0.001  &   0.081$\pm$0.001 \\
        HD306799 &   9.47$\pm$0.12  &   4.66$\pm$0.09  &   4.73$\pm$0.09  &   3.44$\pm$0.09  &   1.35$\pm$0.06  &   1.72$\pm$0.02  &   1.36$\pm$0.06  &   0.174$\pm$0.002  &   0.376$\pm$0.002  &   3.90$\pm$0.07  &   5.52$\pm$0.05  &   0.106$\pm$0.002  &   0.255$\pm$0.002 \\
IRAS\,15060+0947 & ~21.43$\pm$0.31  &   0.66$\pm$0.23  &$-$0.12$\pm$0.28  &$-$1.02$\pm$0.21  &$-$2.38$\pm$0.17  &$-$0.83$\pm$0.13  &   4.03$\pm$0.26  &   0.081$\pm$0.003  &   0.424$\pm$0.005  &  14.34$\pm$0.10  &  10.32$\pm$0.14  &   0.576$\pm$0.003  &   0.972$\pm$0.005 \\
\hline
\end{tabular}
\end{small}
% \end{center}
% \end{table*}
\end{sidewaystable}

\end{document}